\documentclass[sigconf]{acmart}

\setcopyright{acmlicensed}

\newtheorem{theorem}{theorem}

\newtheorem{corollary}{corollary}
\newtheorem*{corollary*}{corollary}

\newenvironment{corollarynum}[1]{%
  \par\smallskip
  \noindent\textbf{Corollary~#1.}\itshape
}{%
  \par\smallskip
}

\newenvironment{proofnum}{%
  \par\smallskip
  \noindent\textbf{Proof}
}{%
  \par\smallskip
}
\usepackage{float}
\usepackage{pifont} 
\usepackage{caption}
\usepackage{subcaption}
\usepackage{booktabs}
\usepackage{listings}
\usepackage{xcolor}
\usepackage[table]{xcolor} 
\newcolumntype{g}{>{\columncolor{green!15}}c}
\usepackage{graphicx}
\usepackage{cleveref}
\usepackage{dsfont}
\usepackage{diagbox} %
\usepackage{multirow}
\usepackage{array}
\usepackage{tabularx}
\usepackage{bm}
\usepackage{empheq}
\usepackage{tikz}
\def\ojoin{\setbox0=\hbox{$\bowtie$}%
  \rule[-.02ex]{.25em}{.4pt}\llap{\rule[\ht0]{.25em}{.4pt}}}

\def\fullouter{\mathbin{\ojoin\mkern-5.8mu\bowtie\mkern-5.8mu\ojoin}}

\usepackage{enumitem}
\setlist[enumerate]{listparindent=\parindent} 
\usepackage{adjustbox}

\acmSubmissionID{}

\begin{document}
\begin{sloppypar}
\title{CUBE: A Cardinality Estimator Based on Neural CDF}


\author{Xiao Yan}
\affiliation{%
  \institution{Northeastern University}
  \city{Shenyang}
  \country{China}}
\email{2201889@stu.neu.edu.cn}

\author{Tiezheng Nie}
\authornote{Corresponding author}
\affiliation{%
  \institution{Northeastern University}
  \city{Shenyang}
  \country{China}}
\email{nietiezheng@mail.neu.edu.cn}

\author{Boyang Fang}
\affiliation{%
  \institution{Northeastern University}
  \city{Shenyang}
  \country{China}}
\email{2110672@stu.neu.edu.cn}

\author{Derong Shen}
\affiliation{%
 \institution{Northeastern University}
 \city{Shenyang}
 \country{China}}
\email{shenderong@cse.neu.edu.cn}

\author{Kou Yue}
\affiliation{%
 \institution{Northeastern University}
 \city{Shenyang}
 \country{China}}
\email{kouyue@cse.neu.edu.cn}

\author{Yu Ge}
\affiliation{%
 \institution{Northeastern University}
 \city{Shenyang}
 \country{China}}
\email{yuge@cse.neu.edu.cn}



\begin{abstract}
Modern database optimizer relies on cardinality estimator, whose accuracy directly affects the optimizer's ability to choose an optimal execution plan. Recent work on data-driven methods has leveraged probabilistic models to achieve higher estimation accuracy, but these approaches cannot guarantee low inference latency at the same time and neglect scalability. As data dimensionality grows, optimization time can even exceed actual query execution time. Furthermore, inference with probabilistic models by sampling or integration procedures unpredictable estimation result and violate stability, which brings unstable performance with query execution and make database tuning hard for database users. In this paper, we propose a novel approach to cardinality estimation based on cumulative distribution function(CDF), which calculates range query cardinality without sampling or integration, ensuring accurate and predictable estimation results. With inference acceleration by merging calculations, we can achieve fast and nearly constant inference speed while maintaining high accuracy, even as dimensionality increases, which is over 10x faster than current state-of-the-art data-driven cardinality estimator. This demonstrates its excellent dimensional scalability, making it well-suited for real-world database applications.

\end{abstract}



\maketitle

\section{Introduction}
Cardinality Estimation (CardEst) is a fundamental problem in database systems, crucial for tasks like query optimization\cite{wang2021we,sun2021learned,han2021cardinality,kim2022learned} and approximate query processing. Query optimizers heavily rely on accurate estimation results to evaluate and select optimal query plans, directly impacting system performance. The errors of traditional CardEst methods can reach up to $10^5$ for complex query workloads and data distributions, leading inefficient query plans. Query-driven methods requires a large amount of workload data for training, obtaining labels by executing queries. Collect such training data is expensive, and the workload data covers a limited set of query patterns and perform poorly on unseen queries. By contrast, data-driven methods are generally more accurate than query-driven approaches and offer strong interpretability, since they can directly capture the underlying probability distribution of table. Therefore, we focus on data-driven method for CardEst problem in this paper.

For accurate CardEst of a given query, it is essential to understand the joint distribution of data from table. This distribution can be effectively characterized using the probability mass function (PMF), probability density Function (PDF), or cumulative distribution function (CDF). In previous works, Naru\cite{yang13deep} and FACE\cite{wang2024face} are typical representatives among the most accurate CardEst methods currently. Naru models relation $T$ as a discrete distribution. Then an autoregressive model is employed to calculate its conditional probabilities $\Pr(x_i | \mathbf x_{<i})$, which can be combined to derive the PMF of relation $T$ as $\Pr(x_1)  \Pr(x_2|x_1) \dots \Pr(x_n|x_1,\dots,x_{d-1})$. Progressive sampling technology is proposed to support range query for Naru. While Face treats relation $T$ as a continuous distribution and applies normalizing flow model for joint probability density estimation $f(\mathbf x)$ directly. Monte Carlo methods is used for high-dimensional integration inference with PDF.


Consider calculating selectivity estimation with joint probability distribution, where Equation ~\eqref{eq:discrete} indicates the probability model of NARU, Equation ~\eqref{eq:continuous} refers to FACE. For $\mathbf x=(x_1,x_2,\ldots, x_d) \in \mathbb{R}^n$, $g(\mathbf x)$ represents its PMF if $\mathbf x$ is a discrete variable and $f(\mathbf x)$ represents the PDF otherwise. 

\vspace{-5mm}
\begin{empheq}[left={\Pr(\mathbf{x} \in \Omega) = \empheqlbrace}]{align}
\sum_{\mathbf{x} \in \Omega} g(\mathbf{x}) & \quad \text{if } \mathbf{x} \text{ is a discrete variable} \label{eq:discrete} \\
\int_{\mathbf{x} \in \Omega} f(\mathbf{x}) \, d\mathbf{x} & \quad \text{if } \mathbf{x} \text{ is a continuous variable} \label{eq:continuous}
\end{empheq}

Constrained by their probability modeling methods for CardEst, they suffer from below drawbacks: (1) negative impacts on both estimation accuracy and inference time, despite efforts for inference optimization. e.g. using the Monte Carlo method\cite{lepage2021adaptive,lepage1978new} to calculate high-dimensional integrals introduces additional estimation errors and extremely high computational time overhead. (2) introduce uncertainty to the inference process, leading to instability and violating the requirement for stable CardEst results.

The CDF can directly provide the cumulative probability for any interval of the random variable, which is convenient for assessing the probability of a variable falling within a specific range. In contrast, determining interval probabilities with a PDF/PMF requires integration/summation, which can be more complex and cause larger error. Therefore, we choose to adopt the joint data cumulative distribution for solving CardEst problem. In summary, the main contributions of our work are as follows:
\begin{itemize}[leftmargin=*]
\item We design \textbf{CUBE}, a \underline{cu}mulative distribution \underline{b}ased cardinality \underline{e}stimator, where the estimation error only comes from the model itself, achieving accurate and stable cardinality estimation results.

\item We propose an inference acceleration strategy for CUBE by merging calculation with model's architecture, resulting a distinct advantage in inference speed and dimensional scalability, which achieves $2^d$ theoretical speedup for a query encoded as a vector with dimensions $d$.

\item We demonstrate the estimation result of CUBE is predictable, which satisfied fundamental rules: monotonicity, validity, consistency, and stability with theoretical proof and experimental evaluation.

\item We propose multiple model training methods for CUBE, support either unsupervised learning with table or supervised learning with workload after data processing including dequantization and normalization.

\item We evaluate CUBE on real-world datasets, which shows that CUBE outperforms representative CardEst methods, achieving higher accuracy with substantially lower inference latency, with a maximum Q-error of less than 3 and inference latency within 1ms on BJAQ and Power datasets, which is over 10x faster than current SOTA data-driven cardinality estimator. Comparative experiments with increasing dimensionality validate the effectiveness of inference acceleration strategy and demonstrate the excellent scalability of CUBE.

\end{itemize}

\section{Preliminaries}
\subsection{Problem Formulation}

Consider a relation $T$ with $d$ attributes $\{ x_1, \cdots, x_d \}$ and an arbitrary query region $\Omega \subseteq R_1 \times \cdots \times R_d$, where $R_i$ denotes the domain of attribute $x_i$ and $T$ can represent a single table or a set of tables $\{T_1, T_2, \cdots, T_n\}$. Given a query $Q$ with $\Omega = \theta_1 \times \cdots \times \theta_d$ representing as below:

\begin{align*}
\text{SELECT COUNT(*) FROM T } \\
\text{ WHERE } \theta_1 \text{ AND } \cdots \text{ AND } \theta_d
\end{align*}

where $\theta_i (i \in [1, k])$ can be an equality predicate like $x = a$, an open range predicate like $x \leq a$, or a close range predicate like $a \leq x \leq b$. Cardinality estimation (CardEst) is to estimate the number of tuples that satisfied all predicate constraints in $Q$ without actually executing the query. In this paper, we discuss an equivalent problem called selectivity estimation, which aims to estimate the percentage of tuples that satisfy the query predicates, represented as follows:

\vspace{-2mm}
\begin{align*}
\text{sel}(\Omega) = \frac{|\sigma_{\mathbf x \in \Omega}|}{|T|}
\end{align*}

where $|T|$ denotes the cardinality of relation $T$
and $|\sigma_{\mathbf x \in \Omega}|$ denotes the number of tuples from $T$ that fall into the region $\Omega$. 

\subsection{Neural Network for CDF Approximation}  
A continuous multivariate CDF approximator (MDMA) is proposed in \cite{gilboa2021mdma}, which learns deep scalar representations for each individual variable and combines them into a tractable yet expressive multivariate CDF. In the following, a capital and lowercase Roman letter (e.g., $F$ and $f$) or Greek letters along with dot above (e.g., $\varphi$ and $\dot{\varphi}$) is used to denote respectively absolutely continuous CDFs of arbitrary dimensions and the corresponding densities. 

\begin{figure}[hbtp]
  \centering
  \includegraphics[width=0.48\textwidth]{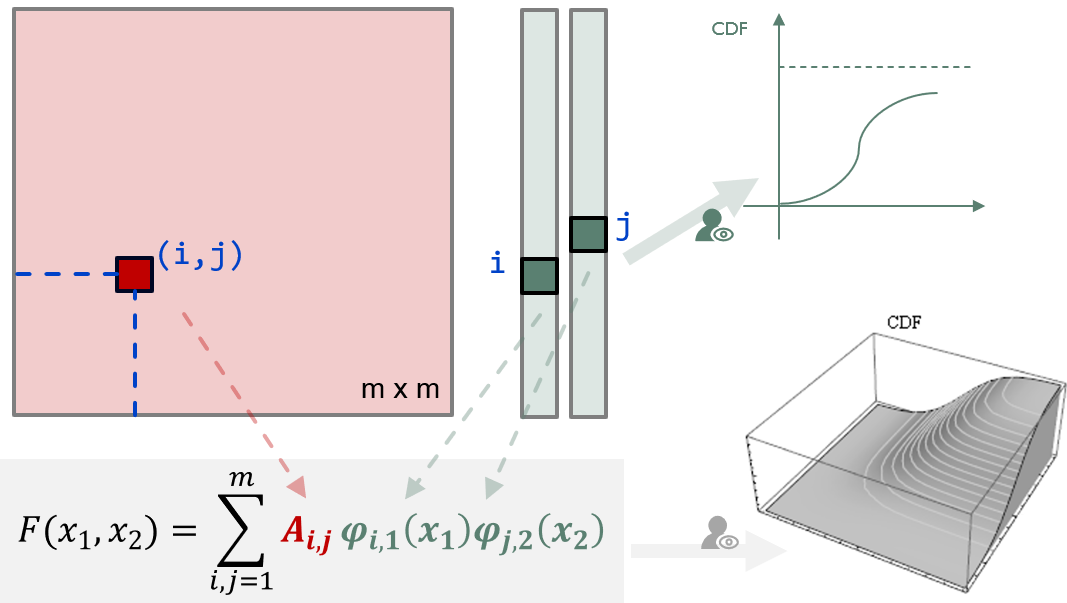}
  \caption{Multivariate CDF obtained by combining univariate CDFs.}
  \label{fig:MDMA}
\end{figure} 

To model joint distributions of $d$ variables on $\mathbb{R}^{d}$, consider a family of univariate CDFs $\{\varphi_{i,j}\,\}_{i\in[m],\,j\in[d]}$ where function $\varphi_{i,j}:\mathbb{R} \rightarrow [0,1]$ satisfy 

\begin{equation}
\begin{aligned}
\lim_{x\to -\infty}\varphi_{i,j}(x) = 0 \,,\\    
\lim_{x\to \infty}\varphi_{i,j}(x) = 1 \,, \\  
\dot{\varphi}_{i,j}(x) = \partial \varphi(x) / \partial x  \geq 0 \,.
\end{aligned}
\end{equation}

Let $A \in \mathbb{R}^{m \times \cdots \times m}$ represents an $m^d$ matrix of elements satisfying

\vspace{-3mm}
\begin{equation}
A_{i_1, \dots, i_d} \ge 0,\, \sum\limits_{i_1, \dots, i_d = 1}^m A_{i_1, \dots, i_d} = 1 .
\end{equation}

We can combine it with the univariate CDFs to obtain a multivariate CDF as below:

\vspace{-3mm}
\begin{equation}
\widehat{F}(\mathbf x) = \sum\limits_{i_{1},\dots,i_{d}=1}^{m}A_{i_{1},\dots,i_{d}}\prod\limits_{j=1}^{d}\varphi_{i_{j},j}(x_{j}).
\end{equation}

The univariate CDF $\varphi(x)$ for $x \in \mathbb{R}$ is modeled using a simple feed-forward network\cite{balle2018variational}

\section{Cardinality Estimation Paradigm under CDF}

In this section, we propose CUBE, a cumulative distribution function based cardinality estimator. We first introduce how CUBE support single-table CardEst using CDF in Section ~\ref{subsec:single_table_cardest} and extend to muti-table CardEst using conditional CDF in Section ~\ref{subsec:multi_table_cardest}. Then we demonstrate how CUBE behaves predictably in Section ~\ref{subsec:predictable_analysis} with formal proofs.

\subsection{Single-Table CardEst Method using CDF} \label{subsec:single_table_cardest}
In our work, we proposed a novel perspective on CardEst problem using CDF with Theorem ~\ref{thm:inference_raw}. This approach minimizes the impact of other aspects on inference as much as possible, despite the model itself. The relation $T$ is considered as a continuous distribution. Since the CDF of data in $T$ can be approximated by our model, $2^d$ appropriate cumulative probabilities are combined to answer an arbitrary query on $d$ dimensional data. 

\begin{theorem}\label{thm:inference_raw}
Given a query region $\Omega = [l_1, u_1] \times \cdots \times [l_d, u_d]$ on a random variable $\mathbf{X} = (X_1, X_2, \cdots, X_d)$ with CDF denoted as $F(\mathbf x) = \Pr(X_1\le x_1,\, X_2\le x_2,\ \ldots,\ X_n\le x_n)$. Let $\mathbf s=(s_1,s_2,\dots,s_d)$  , $\left | s \right | = s_1+s_2+\dots+s_d$ where $s_i \in \{0, 1 \}$. Define $\mathbf v = (v_1, \cdots, v_d)$ with $v_i \in \{l_i, u_i \}$, where $v_i$ takes the value $l_i$ when $s_i=0$ and $u_i$ when $s_i=1$. The probability that $\mathbf x$ fall within $\Omega$ using CDF can be calculated as below:

\begin{equation}
\begin{aligned} \label{eq:inference_raw}
\Pr(\mathbf{x} \in \Omega) &= 
\sum_{s_1\in\{0,1\}}\sum_{s_2\in\{0,1\}}\cdots \sum_{s_n\in\{0,1\}}(-1)^{n-\sum_{j=1}^n s_j}\ F(\mathbf{v})  \\
&= \sum_{\mathbf{s} \in \{0,1\}^n} (-1)^{n - |\mathbf{s}|} \ F(\mathbf{v}) 
\end{aligned}
\end{equation}

\end{theorem}

\begin{figure}[!h]
  \centering
  \includegraphics[width=0.34\textwidth]{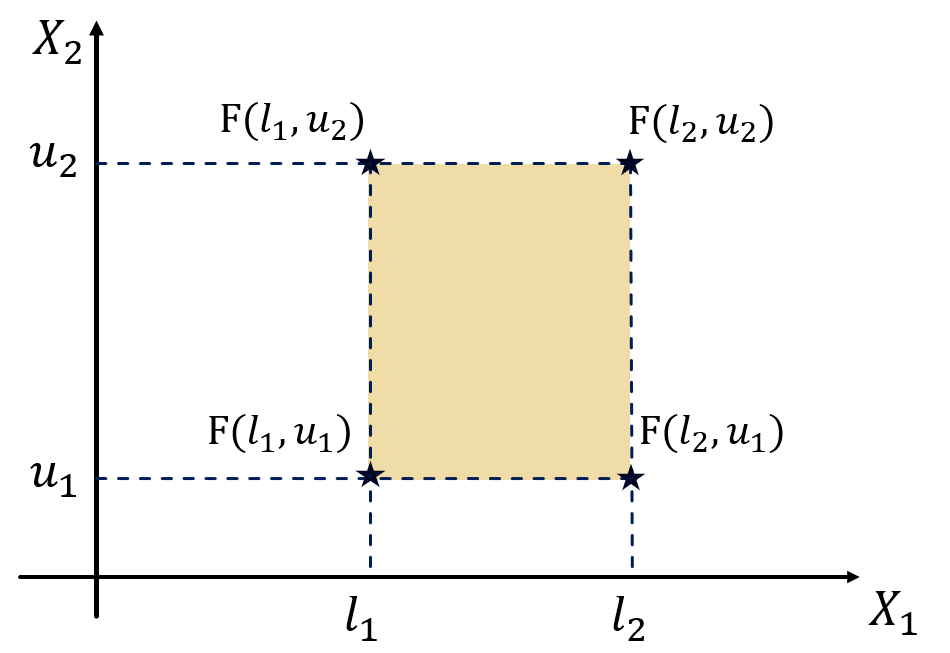}
  \caption{CardEst with CDF on 2-dimensional data}
  \label{fig:inference_raw_vis}
\end{figure}

As shown in Figure~\ref{fig:inference_raw_vis}, when d = 2, selectivity with query region $\Omega = (l_1, u_1] \times (l_2, u_2]$ can be obtained as 
\begin{equation}
\Pr(\mathbf x \in \Omega)=F(u_1, u_2) - F(u_1, l_2) - F(l_1, u_2) + F(l_1, l_2)
\end{equation}

\vspace{-4mm}

\vspace{-6mm}
\subsubsection{\textbf{Continuity Correction}}  \mbox{}\\ \label{subsec:continuity_correction}

\vspace{-2mm}
The distribution of data from relation $T$ is discrete, which can be approximated with a continuous distribution learned by our model.  It might cause estimation error if we directly feed the query into model. Thus, a \textbf{continuity correction factor} should be added properly to fixed that issue. Let $\widehat{F}(\mathbf{x})$ represent the CDF of the data derived from relation $T$ using MDMA. For example, we can simply approximate the probability by combining two cumulative probabilities as $\Pr(a \leq x \leq b) = \sum_{x=a}^{b} g(x) \approx \widehat{F}(b) - \widehat{F}(a)$. However, this simple approximation has the following shortcoming: although $\Pr(X \geq a)$ and $\Pr(X>a)$ are always equal for the continuous distribution, these probabilities typically have different values for the discrete distribution which may cause large estimation error.

A continuity correction is an adjustment that is made when a discrete distribution is approximated by a continuous distribution. Define the continuity correction factor as $\mathbf w = (w_1, \cdots, w_d)$, where $w_i$ is the data precision along dimension~$i$ (e.g., if a column is integer-valued, $w_i=1$). Then, the probability of a single data point $\mathbf x$ can be approximated using continuity correction as $\Pr(\mathbf X = \mathbf x) \approx  \widehat{F}(\mathbf x+\mathbf w )-\widehat{F}(\mathbf x)$. Similarly, cumulative probability after correction is $\Pr(\mathbf X \leq \mathbf x) \approx  \widehat{F}(\mathbf x+\mathbf w )$

After we convert different types of queries into range queries, continuity correction with a proper factor should be done in the right place. In the case of a range interval with a left-open or right-close endpoint, the left endpoint or the right endpoint would be adjusted by adding the factor, respectively, while the other endpoints remain unchanged. For example, given a query predicate like $a \leq x \leq b$ with closed range interval, then the range interval should be adjusted as $[a, b+w)$  under continuous distribution. Other situations can be handled correctly by our continuous model without continuity correction.

\begin{figure}[htbp]
    \centering
    \includegraphics[width=0.4\textwidth]{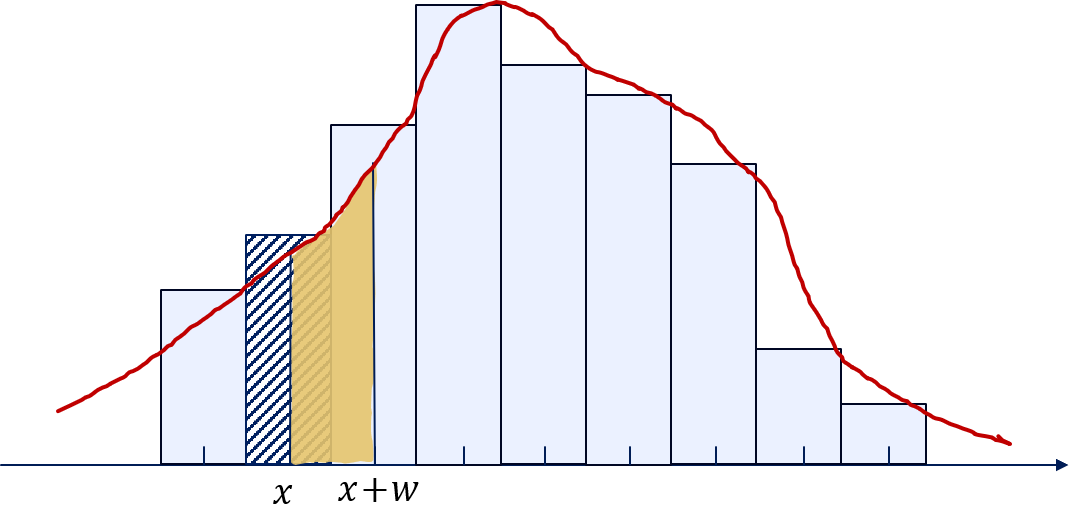}
    \caption{continuity correction on 1-dimensional data.}
\end{figure}

\subsubsection{\textbf{Inference Acceleration}} \mbox{}\\
Since the naive inference method for CardEst only performs well under low dimensions, the calculation time increases explosively as the data dimension increases. Thus, we propose an optimization strategy to improve model's inference efficiency by merging calculation, which makes our model practical under high-dimensional data and additionally improves the efficiency of supervised training.

According to Equation\eqref{eq:inference_raw}, answer a query encoded as a vector with dimensions $d$, a total of $2^d$ CDF value computations are required. When d is not too large, this problem is not very pronounced, and compared to method such as Monte Carlo integration for high-dimensional data, our computational approach can be considerably faster. However, as d increases, an exponential explosion phenomenon occurs. For example, when d reaches 30, we need approximately $10^9$ CDF calculations to obtain the final estimation result. Consequently, the model proposed in this paper no longer holds a significant advantage in terms of inference speed and becomes unusable. 

Next, we would discuss how we use some strategies to speed up model inference and make model available even under extremely high-dimensional data.

\underline{\textbf{Strategy 1: Pre-calculation}} Since our model computes the multivariate CDF by combining multiple individual CDF models, for the variables $(x_1, x_2, \ldots, x_n)$ involved in $F_X(x_1, x_2, \ldots, x_n)$, each $x_j$ (where $j \in {1, 2, \ldots, n}$) is only evaluated within its corresponding individual CDF model and contributes to the final multivariate CDF. Therefore, we can pre-calculate the contribution values of all $x_j$ to the overall estimation, denoted as $\{\varphi_{i,j}\}_{i\in[m],{j\in[d]}}$. This can be considered as a caching strategy, where when the overall CDF estimation is needed, the cached individual CDFs are weighted and aggregated accordingly. Compared to the previous requirement of calculating $2^n$ individual variable CDFs,  now $2 * n$ times computation is enough.

However, when using the above-mentioned strategie for model inference, the acceleration effect is very limited. The time required for the weighted summation of individual variable CDFs is significantly greater than the time needed to compute individual variable CDFs. Consequently, the optimizations we have implemented have had only a marginal effect on the overall inference time. Thus, we need a more comprehensive optimization approach.

\underline{\textbf{Strategy 2: Merged-Calculation}}  Considering the characteristics of our model, we found a concise and efficient approach which can combines $2^n$ computations for CardEst inference in Theorem ~\ref{thm:inference_raw} and fundamentally enhance inference speed through mathematical derivations, which can be formulated as corollary ~\ref{corollary:inference_optimized} below.

\begin{corollary} \label{corollary:inference_optimized}
Given random variable $\mathbf{X} = (X_1, X_2, \dots, X_d)$ with CDF approximation as $\sum\limits_{i_{1},\dots,i_{d}=1}^{m}A_{i_{1},\dots,i_{d}}\prod\limits_{j=1}^{d}\varphi_{i_{j},j}(x_{j})$ and a query region $\Omega = [l_1, u_1] \times \cdots \times [l_d, u_d]$. The probability that $\mathbf{x}$ falls in $\Omega$ is
\begin{equation}\label{eq:inference_optimized}
\begin{split}
\Pr(\mathbf x \in \Omega)
&= \sum_{\mathbf{s}\in\{0,1\}^n} (-1)^{n-|\mathbf{s}|}\, F(\mathbf{v})\\
&\approx
    \sum_{i_{1},\dots,i_{d}=1}^{m}
      A_{i_{1},\dots,i_{d}}
      \prod_{j=1}^{d}
        \bigl[\varphi_{i_{j},j}(u_{j}) - \varphi_{i_{j},j}(l_{j})\bigr].
\end{split}
\end{equation}
\end{corollary}

\vspace{-4mm}
\begin{figure}[ht]
\centering
\begin{subfigure}[b]{0.48\textwidth}    
    \centering
    \includegraphics[width=\textwidth]{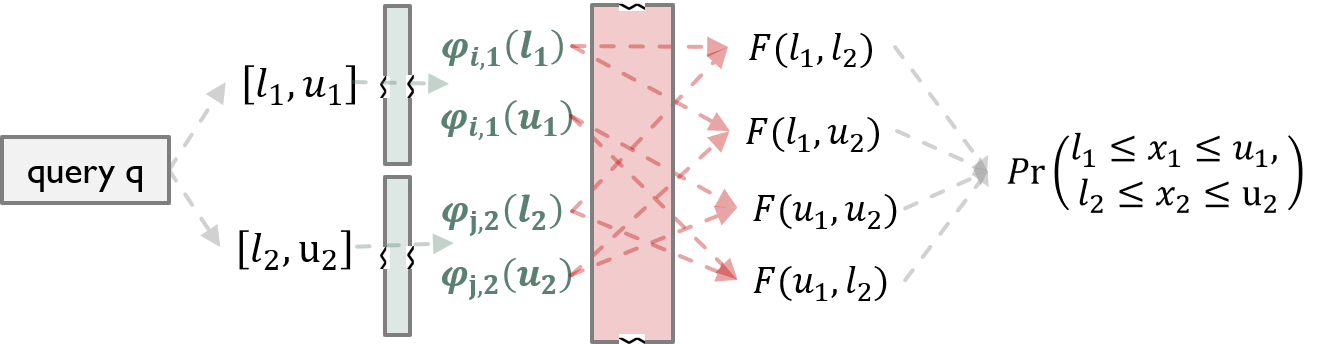}
    \caption{}
    \label{fig:inference_raw}
\end{subfigure}%
\hfill
\begin{subfigure}[b]{0.48\textwidth}   
    \centering
    \includegraphics[width=\textwidth]{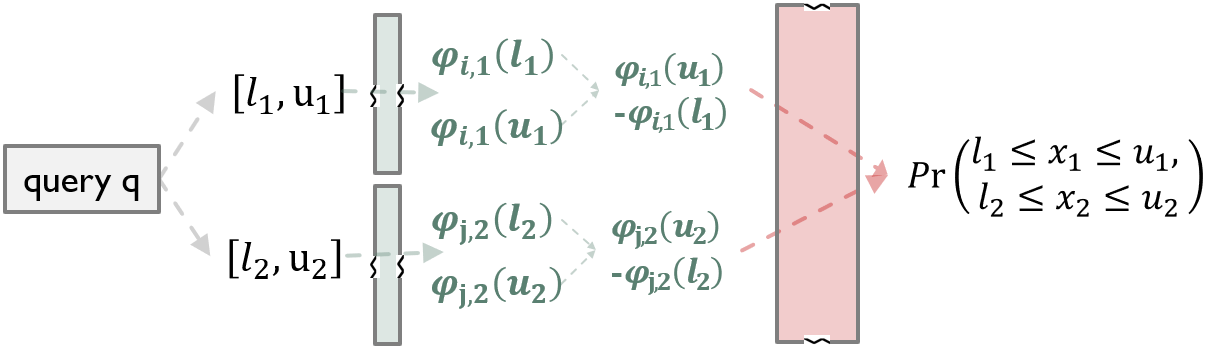}
    \caption{}
    \label{fig:inference_optimized}
\end{subfigure}
\caption{(a) inference procedure before optimization; (b) inference procedure after merged-calculation.}
\label{fig:inference_procedure}
\end{figure} 

\vspace{-2mm}
Therefore, we no longer need to calculate $2^n$  joint cumulative probabilities for answering an arbitrary query, once is enough now. We show the details when d = 2 in Figure ~\ref{fig:inference_procedure}.

\noindent{\textbf{Theoretical Speedup}} 
We define the theoretical speedup as the ratio of time complexity between the raw inference strategy and the optimized inference strategy. By analyzing the mathematical formulation of the model inference, \textbf{the theoretical speedup is $2^d$ for a query encoded as a vector with dimensions $d$.} Apparently, the higher the dimension of the data, the more pronounced the acceleration effect. This greatly enhances the applicability of the model to high-dimensional data.

\vspace{-2mm}
\subsection{Muti-Table CardEst Method using Conditional CDF} \label{subsec:multi_table_cardest}

We build a global cardinality estimator based on join schema to efficiently support inference for multi-table join queries. Initially, we introduce the method of using a global cardinality estimator for multi-table join estimation, allowing a single model to support various queries. Subsequently, we demonstrate conditional CDF can be accurately and efficiently utilized to compute the probability expectation under global schema, which equals performing inference procedure for muti-table CardEst.

\begin{figure}[h]
  \centering
  \includegraphics[width=0.25\textwidth]{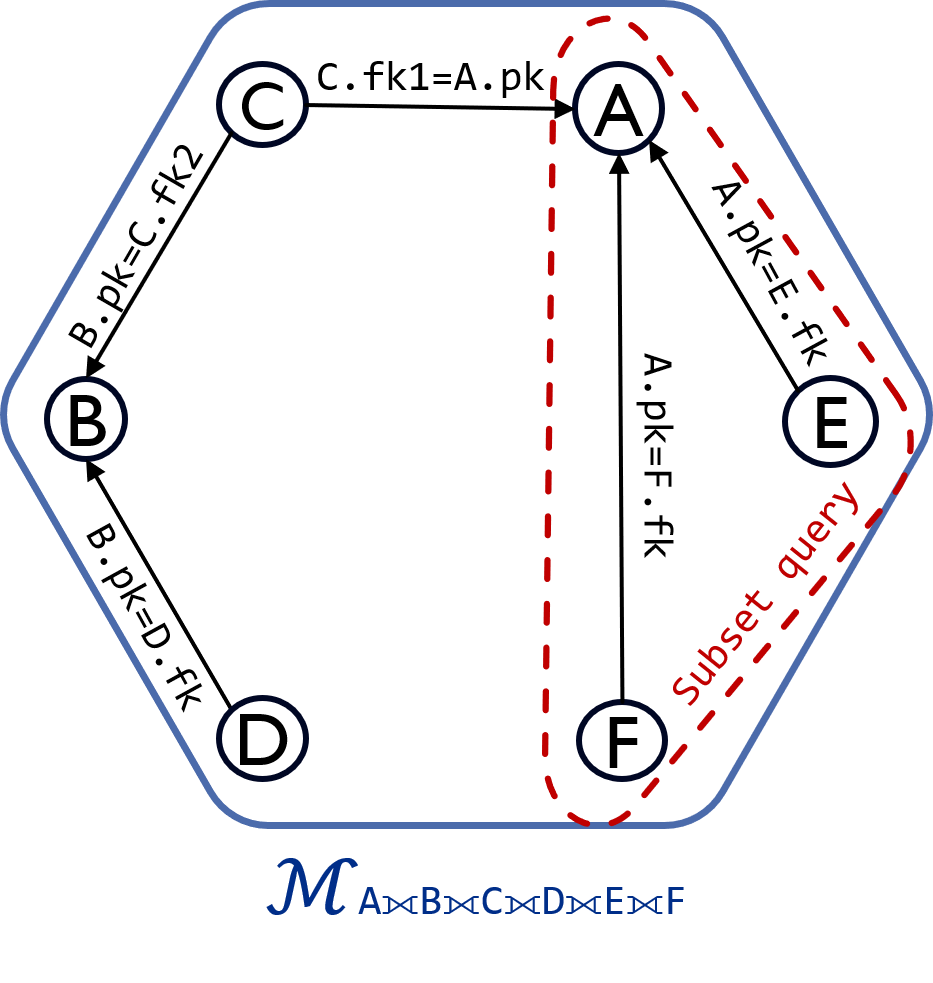}
  \caption{Global model for multi-table CardEst, building on full outer join of table $A,B,C,D,E,F$ with subset query involving table $A,E,F$} 
  \label{fig:global_model}
\end{figure}

\begin{figure}[htbp]
\captionsetup[subfigure]{justification=centering}
    \centering
    \begin{subfigure}[b]{0.45\columnwidth}
    \begin{tikzpicture}[inner sep=1pt, text centered]
      \node {
        \tabcolsep=0.06cm
        \begin{tabular}{@{} c c @{}}
          $A.pk$ & $A.x$ \\ \midrule
          1 & 0.7 \\
          2 & 0.3 \\
          3 & 0.5
          \end{tabular}
          \quad
          \begin{tabular}{@{} c c @{}}
            $E.pk$ & $E.x$ \\ \midrule
            2 & 10 \\
            2 & 40 \\
            3 & 20
          \end{tabular}
      };
    \end{tikzpicture}
    \caption{}
    \label{fig:ce_join_example_base_tables}
    \end{subfigure}
    \hspace{-1em}%
    \begin{subfigure}[b]{0.45\columnwidth}
    \begin{tikzpicture}[inner sep=1pt, text centered]
      \node {
        \tabcolsep=0.06cm
        \begin{tabular}{@{} c c l c l l lll @{}}
          $A.pk$ & $A.x$ & $\textcolor{blue}{\mathcal{F}_{A.pk}}$ &  $E.x$  & $\textcolor{blue}{\mathcal{F}_{E.fk}}$ & $\textcolor{orange}{\mathds{1}_A}$  & $\textcolor{orange}{\mathds{1}_E}$ \\ \midrule
        1 & 0.7   & \textcolor{blue}{1} & $\varnothing$  & \textcolor{blue}{1} & \textcolor{orange}{1} & \textcolor{orange}{0} \\
        2 & 0.3   & \textcolor{blue}{1} & 10  & \textcolor{blue}{2} & \textcolor{orange}{1} & \textcolor{orange}{1} \\
        2 & 0.3   & \textcolor{blue}{1} & 40  & \textcolor{blue}{2}    & \textcolor{orange}{1} & \textcolor{orange}{1} \\
        3 & 0.5   & \textcolor{blue}{1} & 20  & \textcolor{blue}{1}     & \textcolor{orange}{1} & \textcolor{orange}{1}
      \end{tabular}
      };
    \end{tikzpicture}
    \caption{}
    \label{fig:ce_join_example_full} 
    \end{subfigure}
    \caption{Multi-table CardEst example: (a) Base table with constraint $A.pk\leftarrow E.fk$; (b) Full outer join table obtained by table $A$ and $E$ along with fanout column $\mathcal{F}$ and indicator column $\mathds{1}$} 
    \label{fig:ce_join_example}
\end{figure}

\subsubsection{\textbf{Global Model based on Join Scheme}} \mbox{} \\
We employ a global cardinality estimator for multi-table join estimation as shown in Figure ~\ref{fig:global_model}. Specifically, we can perform a full outer join across all involved tables in queries, resulting in a completely flattened wide table. A global model is learned based on this wide table, supporting inference for any subset query within the join schema, mainly because of the following advantages: 
\begin{enumerate}
\item \textbf{Simplicity.} Using multiple estimators, each covering only specific join templates (table subsets), becomes hard to scale with an exponential increase in possible join templates as the number of tables grows. Additionally, training a single estimator is easier than training multiple estimators.
\item \textbf{Accuracy.} Employing multiple estimators may reduce estimation accuracy. If a query's table subset is not directly covered by a single estimator, combining results from multiple estimators typically requires independence assumptions. Violations of these assumptions degrade estimation accuracy.
\end{enumerate}

Since the global model learns joint probability distribution from full outer join data, querying the global model by default returns estimation results from entire outer join. Thus, additional processing is required for inner join queries and subset queries, discussed as below:

\begin{itemize}[leftmargin=*]
\item \textbf{Inner Join Queries.} The full outer join includes tuples from base tables without matching join records. To support inner join queries, non-matching tuples must be removed. We fill mismatching records with values outside the domain of predicates during model training, and restrict the query predicates within their valid domains during inference procedure to enable correct estimation. \\[-3mm]

An indicator column $\mathds{1}_T$ is added to mark matching tuples for table $T$, enabling correct estimation. For example, in the full outer join, the result for $E.x \neq 40$ is 3, but the actual value in the original table $E$ is 2. Specifically, we add an extra indicator column $\mathds{1}_T$ for full outer join to identify whether a joined tuple has a matching record in table $T$. If a tuple in the full outer join has a valid matching record in table $T$, then $\mathds{1}_T$ is set to 1; otherwise, it is set to 0, as shown in Figure~\ref{fig:ce_join_example}. For inner join table $|A \Join E|$ the estimation result with query $E.x \neq 40$ is as follows:
\begin{align*}
|A \Join E|_{E.x \neq 40}
&= |A \fullouter  E|\cdot\mathbb{E}\left(\mathds{1}_{E.x \neq 40}\cdot \mathds{1}_E \right)\\ 
&= 5\cdot\frac{1+1}{5}=2
\end{align*}

\item \textbf{Subset Queries.} Duplicate join key values inflate cardinality counts in full outer joins. To correct this, a \textbf{fanout factor} column $\mathcal{F}_{T.key}$ is added for each join key, representing the occurrence count, allowing accurate down-scaling adjustments\cite{hilprecht13deepdb,yang2020neurocard}. \\[-2mm]

For example, in the full outer join, the result for $A.x < 0.6$ is 3, but the actual result in the original table $A$ is 2. This is because $A.pk=2$ appears twice in $E.fk$, leading to double counting. Therefore, the estimated result needs to be down-scaled. To address this issue, for each join key column $T.key$, an additional column $\mathcal{F}_{T.key}$ is added to the full outer join as the \textbf{fanout factor}, used for subsequent down-scaling. It is defined as the number of times each key value appears in $T.key$. For example, in Table~\ref{fig:ce_join_example_full}, the value 2 appears twice in $E.fk$, so its fanout is $\mathcal{F}_{E.x}(2)=2$, indicating that there are two records in table $E$ satisfying the join key $join\_key=2$. It is important to note that for each pair of related tables, the fanout factor needs to be computed only once, usually during the construction of the full outer join data, and should be maintained promptly during data updates. For table $A$, the estimation result for query $A.x \geq 0.2$ is as follows:
\begin{align*}
|A|_{A.x < 0.6 }
&= |A \fullouter  E|\cdot\mathbb{E}\left(\frac{1}{\mathcal{F}_{E.fk}} \cdot \mathds{1}_{A.x < 0.6}\right)\\ 
&= 5\cdot\frac{1/2+1/2+1}{5}=2
\end{align*}

\end{itemize}

\subsubsection{\textbf{Inference with Probability Expectation}} \mbox{} \\

We formally define the inference method as Theorem~\ref{thm:ce_join} for muti-table CardEst under a global schema, which can be seen as calculating probability expectation essentially.

\begin{theorem} \label{thm:ce_join}
Let $J$ represent all tables, $Q$ be the subset participating in the query, and $O = J \setminus Q$ be non-participating tables. With a global schema as a directed acyclic graph, for each $T \in Q$, there is a unique path from each $R \in O$ to $T$, and the join key $R.key$ on this path is the key for correction. For each $R \in O$, the fanout factor $\mathcal{F}_{R.key}$ is uniquely determined. Given query predicate domain $\Omega$, the query result is:

\begin{align}
|J|\cdot\mathbb{E}\left(\frac{1}{\prod_{R \in O}\mathcal{F}_{R.key}}\cdot\mathds{1}_{X \in \Omega}\right)
\end{align}

\end{theorem}

Since MDMA is continuously differentiable, probability expectation for multi-table join CardEst can be computed by integrating the PDF at any point $\bm{x}$ within the domain $\Omega$. However, due to the curse of dimensionality, direct numerical integration in high-dimensional spaces is often infeasible. Therefore, Monte Carlo methods are commonly used to approximate high-dimensional integrals. Nevertheless, such approaches suffer significant drawbacks: as dimensionality increases, the size of the integration domain grows, requiring more sample points, thus consuming more computational resources and amplifying errors.

To address this, Corollary~\ref{corollary:inference_expectation} is proposed to accurately compute the probability expectation using conditional CDF, merging intermediate computation steps to efficiently infer CardEst results. 

\vspace{-6mm}
\begin{figure}[h]
  \centering
  \includegraphics[width=0.35\textwidth]{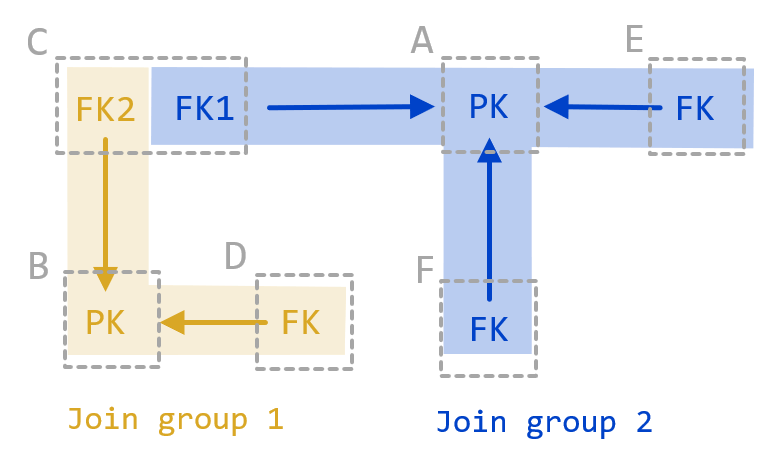}
  \caption{join groups under foreign key constraint} 
  \vspace{-2mm}
  \label{fig:join_key_group}
\end{figure}

We observe that both sides of an equi-join relation represent the semantically equivalent join keys. Thus, we divided them into the same \textbf{join group}. In the full outer join, there exist multiple such join groups, denoted as $X^{(1)} = (X_1, \ldots, X_k)$, where $X_i$ represents the equivalent key variable in i-th join group. Under foreign key constraint, the value of a join group depends on the primary key of the associated base table. For example, Figure~\ref{fig:join_key_group} shows the primary-foreign key relations among tables $A, B, C, D, E, F$, where the join groups are $X^{(1)} = (A.pk, B.pk)$. The fanout columns can thus be expressed as functions of the join groups, denoted as $W_{fanout}(X^{(1)})$. Ultimately, the multi-table CardEst result can be derived from Corollary~\ref{corollary:inference_expectation}.

\begin{corollary} \label{corollary:inference_expectation}
Given random variable $X = (X^{(1)}, X^{(2)})$ with CDF approximation as
$\sum\limits_{i_{1},\dots,i_{d}=1}^{m}A_{i_{1},\dots,i_{d}}\prod\limits_{j=1}^{d}\varphi_{i_{j},j}(x_{j})$, where random vectors $X^{(1)} = (X_1, \ldots, X_k)$ and $X^{(2)} = (X_{k+1}, \ldots, X_d)$ represent two sub-vectors of $X$. Let $W(X^{(1)})$ be a function depending only on sub-vector $X^{(1)}$, with domain $R_{X}^{(1)}$ for $X^{(1)}$ and domain $\Omega = [l_{k+1}, u_{k+1}] \times \dots \times [l_{d}, u_{d}]$ for $X^{(2)}$. Then the expectation of $W(X^{(1)})$ within region $\Omega$ is present as:

\begin{equation}
\begin{aligned} 
&\mathbb{E}\left[\mathds{1}_{\{x^{(2)} \in \Omega\}}\cdot W(x^{(1)})\right] \\
&= \sum_{i_{1},\dots,i_{d}=1}^{m}A_{i_{1},\dots,i_{d}}\left\{\prod\limits_{j=k+1}^{d}[\varphi_{i_{j},j}(u_{j})-\varphi_{i_{j},j}(l_{j})]\right\}\cdot G_{i_1,\dots,i_k}(x^{(1)}).
\end{aligned}
\end{equation}

where intermediate function $G_{i_1,\dots,i_k}(x^{(1)})$ is defined as:

\begin{equation}
\begin{aligned} 
&G_{i_1,\dots,i_k}(x^{(1)}) \\
&= \sum_{x^{(1)} \in R_{X}^{(1)}}\frac{\prod\limits_{j=1}^{k}\dot{\varphi}_{i_{j},j}(x_{j})}{\sum\limits_{i_{1},\dots,i_{k}=1}^{m}A_{i_{1},\dots,i_{k}}\prod\limits_{j=1}^{k}\dot{\varphi}_{i_{j},j}(x_{j})}\cdot\Pr\left(X^{(1)}=x^{(1)}\right)\cdot W(x^{(1)}).
\end{aligned}
\end{equation}

\end{corollary}

Note that when the number of join groups is large, the value combinations of equivalent key variables can be numerous, making it computationally inefficient to enumerate all possible values of $X^{(1)}$. However, join key values in the full outer join are constrained, meaning not all combinations of $x^{(1)}$ are valid. For example, under the schema illustrated in Figure~\ref{fig:join_key_group}, the actual values of join groups $X^{(1)} = (A.pk, B.pk)$ are constrained by foreign key pairs $(C.fk1, C.fk2)$ in table $C$. Thus, the valid domain of join groups $R_{X}^{(1)}$ and corresponding frequency statistics $\text{freq}_{X^{(1)}}$ can be precomputed during the full table join. These precomputed values facilitate efficient inference during subsequent queries.

\subsection{Predictable Analysis} \label{subsec:predictable_analysis}
In database query optimization, the results of CardEst should be predictable~\cite{wang2021we}, satisfying four fundamental rules: \textbf{monotonicity}, \textbf{validity}, \textbf{consistency}, and \textbf{stability} when data remains unchanged. Violation of these logical rules may confuse database developers and users and negatively impact query performance. We demonstrate that CUBE satisfies these rules through analysis and proofs 

\subsubsection{\textbf{Monotonicity}} \mbox{} \\
When the query predicates become stricter, the estimation result should not increase; conversely, estimation result should not decrease when the query predicates become looser. The corresponding formal description and proof are shown below.

\begin{theorem} \label{thm:monotonicity}
Given random variable $\mathbf{X} = (X_1, X_2, \dots, X_d)$ with domain $\Omega = [l_1, u_1] \times \dots \times [l_d, u_d]$, and domain variations $\boldsymbol{\varepsilon} = (\varepsilon_1, \varepsilon_2, \dots, \varepsilon_d)$, for the expanded domain $\Omega_{\boldsymbol{\varepsilon}^{+}} = [l_1, u_1+\varepsilon_1] \times \cdots \times [l_d, u_d+\varepsilon_d]$ and the narrowed domain $\Omega_{\boldsymbol{\varepsilon}^{-}} = [l_1, u_1+\varepsilon_1] \times \cdots \times [l_d, u_d+\varepsilon_d]$, we have:

\begin{equation}
\begin{aligned}
\Pr(\mathbf x \in \Omega_{\boldsymbol{\varepsilon}^{+}}) \geq \Pr(\mathbf x \in \Omega), \\ 
\Pr(\mathbf x \in \Omega_{\boldsymbol{\varepsilon}^{-}}) \leq \Pr(\mathbf x \in \Omega).
\end{aligned}
\end{equation}

\end{theorem}

\begin{proof} \label{proof:monotonicity}
For the expanded domain $\Omega_{\boldsymbol{\varepsilon}^{+}}$, according to Corollary~\ref{corollary:inference_optimized}, the probability of $\mathbf{x}$ falling within $\Omega_{\boldsymbol{\varepsilon}^{+}}$ is:
\begin{align*}
\Pr(\mathbf x \in \Omega_{\boldsymbol{\varepsilon}^{+}})
&= \sum\limits_{i_{1},\dots,i_{d}=1}^{m} A_{i_{1},\dots,i_{d}} \prod\limits_{j=1}^{d} \Bigl[\varphi_{i_{j},j}(u_{j}+\varepsilon_j) - \varphi_{i_{j},j}(l_{j})\Bigr].
\end{align*}

Due to the monotonicity of the single-variable cumulative distribution function $\varphi_{i_{j},j}(x_j)$:
\begin{align*}
\varphi_{i_{j},j}(u_{j}+\varepsilon_j) \geq \varphi_{i_{j},j}(u_{j}),
\end{align*}
we have:
\begin{equation*}
\varphi_{i_{j},j}(u_{j}+\varepsilon_j) - \varphi_{i_{j},j}(l_{j}) \geq \varphi_{i_{j},j}(u_{j}) - \varphi_{i_{j},j}(l_{j}),
\end{equation*}
thus clearly:
\begin{equation*}
\Pr(\mathbf x \in \Omega_{\boldsymbol{\varepsilon}^{+}}) \geq \Pr(\mathbf x \in \Omega).
\end{equation*}
The proof for the narrower domain follows similarly.
\end{proof}

\subsubsection{\textbf{Validity}} \mbox{} \\
For invalid predicates, CardEst result should be zero, for example:
\begin{align*}
\text{SELECT COUNT(*) FROM T WHERE} \quad 10 \leq x_i \leq 1.
\end{align*}

The formal description and proof are shown as below:

\begin{theorem} \label{thm:validity}
Given random variable $\mathbf{X} = (X_1, X_2, \dots, X_d)$ with domain $\Omega = [l_1, u_1] \times \dots \times [l_d, u_d]$, if $\exists k$ satisfying $l_k > u_k$ and $\forall j \neq k$ having $l_j < u_j$, the probability of $\mathbf{x}$ falling within $\Omega$ is:
\begin{equation}
\Pr(\mathbf x \in \Omega) = 0.
\end{equation}
\end{theorem}

\begin{proof} \label{proof:validity}
From the monotonicity of univariate CDF:
\begin{align*}
\varphi_{i_{k},k}(u_{k}) - \varphi_{i_{k},k}(l_{k}) \leq 0, \quad j=k; \\ 
\varphi_{i_{j},j}(u_{j}) - \varphi_{i_{j},j}(l_{j}) \geq 0, \quad j \neq k.
\end{align*}
Applying a clipping function to eliminate negatives ensures validity:
\begin{align*}
\Pr(\mathbf x \in \Omega) &= \sum\limits_{i_{1},\dots,i_{d}=1}^{m} A_{i_{1},\dots,i_{d}} \prod\limits_{j=1}^{d} \max\{\varphi_{i_{j},j}(u_{j}) - \varphi_{i_{j},j}(l_{j}), 0\} = 0.
\end{align*}
\end{proof}

\subsubsection{\textbf{Consistency}} \mbox{} \\
The estimated cardinality of a query should be equal to the sum of the estimated cardinalities of its decomposed sub-queries. The formal description and proof are shown as below: 

\begin{theorem} \label{thm:consistency}
Given random variable $\mathbf{X} = (X_1, X_2, \dots, X_d)$ with domain $\Omega = [l_1, u_1] \times \dots \times [l_d, u_d]$, for two subdomains $\Omega_1 = [l_1, u_1] \dots \times [l_k, m_k] \times \dots \times [l_d, u_d]$ and $\Omega_2 = [l_1, u_1] \dots \times [m_k, u_k] \times \dots \times [l_d, u_d]$ satisfying $\Omega = \Omega_1 + \Omega_2$, the probability that $\mathbf{x}$ falls into region $\Omega$ satisfies:

\begin{equation}
\Pr(\mathbf x \in \Omega) = \Pr(\mathbf x \in \Omega_1) + \Pr(\mathbf x \in \Omega_2)
\end{equation}
\end{theorem}

\begin{proof} \label{proof:consistency}

According to Corollary~\ref{corollary:inference_optimized}, the sum of the probabilities that $\mathbf{x}$ falls into regions $\Omega_1$ and $\Omega_2$ is:

\begin{align*}
&\Pr(\mathbf x\in\Omega_1)+\Pr(\mathbf x\in\Omega_2)\\
&=\sum_{i_1,\dots,i_d=1}^m
A_{i_1,\dots,i_d}\,
\prod_{\substack{j=1\\j\neq k}}^{d}
[\varphi_{i_j,j}(u_j)-\varphi_{i_j,j}(l_j)] 
\times [\varphi_{i_k,k}(m_k)-\varphi_{i_k,k}(l_k)]\\
& \quad +\sum_{i_1,\dots,i_d=1}^m
A_{i_1,\dots,i_d}\,
\prod_{\substack{j=1\\j\neq k}}^{d}
[\varphi_{i_j,j}(u_j)-\varphi_{i_j,j}(l_j)]
\times [\varphi_{i_k,k}(u_k)-\varphi_{i_k,k}(m_k)]\\
&=\sum_{i_1,\dots,i_d=1}^m
A_{i_1,\dots,i_d}\,
\prod_{j=1}^{d}
[\varphi_{i_j,j}(u_j)-\varphi_{i_j,j}(l_j)]\\
&=\Pr(\mathbf x\in\Omega)\,.
\end{align*}

\end{proof}

\subsubsection{\textbf{Stability}} \mbox{} \\
The estimation result for the same query must remain consistent without fluctuation. Since the parameters of CUBE are fixed after training and the inference result is uniquely determined for a given query input (by Corollaries~\ref{corollary:inference_optimized} and \ref{corollary:inference_expectation}), CUBE inherently produces stable estimates results.

\section{Model Training}

In this section, we illustrate how CUBE fits the joint distribution using either table through unsupervised learning or workload through supervised learning after data preprocessing, which can help our model learn the cumulative distribution effectively.

\begin{figure}[hbtp]
  \includegraphics[width=0.54\textwidth]{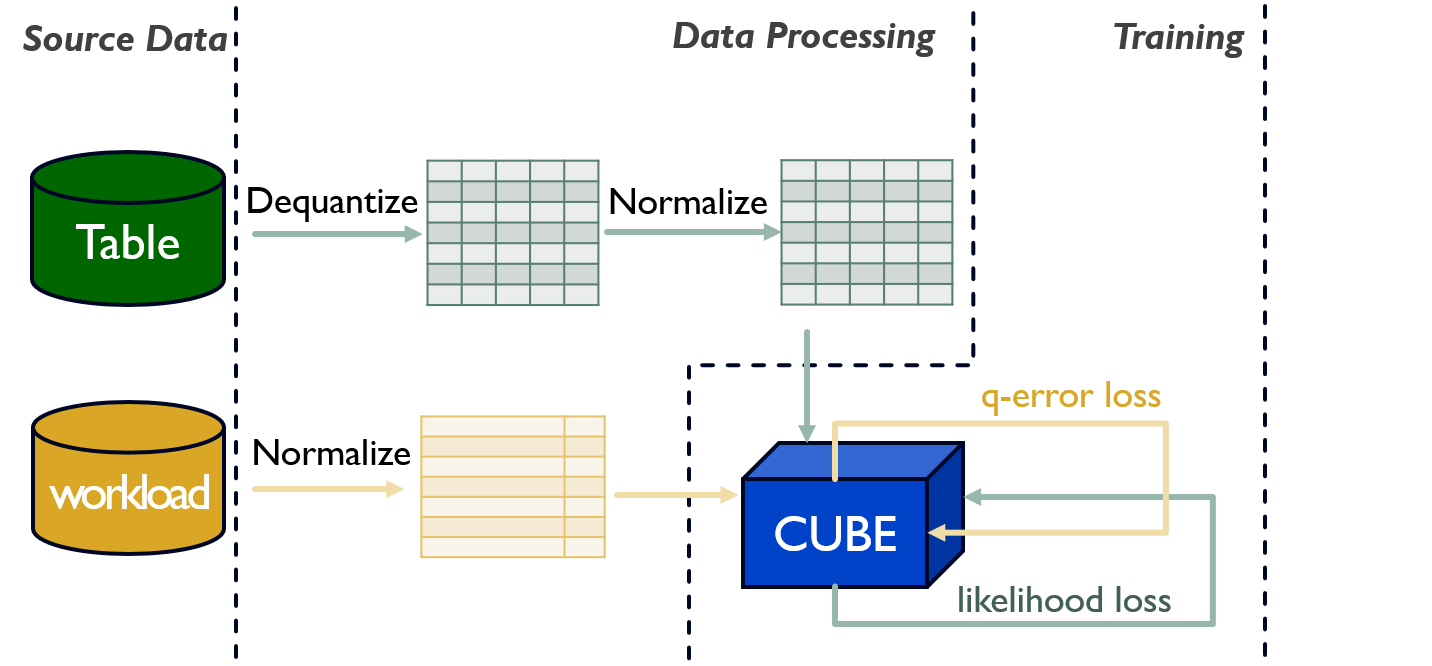}
  \caption{Training framework of CUBE} 
  \label{fig:framework}
\end{figure}


\subsection{\textbf{Data preprocessing}} \mbox{} \\

\vspace{-8mm}
\noindent \textbf{Supported data types} Numerical and categorical data are commonly used in databases. Therefore, we support them with preprocessing, while the others can be added in an extended manner. Since categorical data always appears in different forms as numbers, Greek letters, strings or others, we convert them into numerical datas. For example, a library database has a attribute $Cate$ with books that might be classified as Sciences, Philosophy, Arts, Geography, and History. These categories can be regarded as numerical values in the set $\{1, 2, 3, 4, 5\}$.

\noindent \textbf{Dequantization} 
Modeling discrete data using a continuous distribution can lead to arbitrarily high density values, which locates narrow high-density spikes on discrete values. However, the continuous model does not necessarily approximate a probability mass because the spaces of discrete and continuous variables are topologically different, which is that the density of a data point is zero under continuous model. To deal with this issue, right amount of uniform noise is added which dequantizes the data. By adopting this approach, the log-likelihood of the continuous model on the dequantized data is closely related to the log-likelihood of a discrete model on the discrete data which improves the model's performance\cite{theis2016note}. 

\begin{figure}[h]
    \centering
    \begin{subfigure}[b]{0.24\textwidth}
      \includegraphics[width=\textwidth]{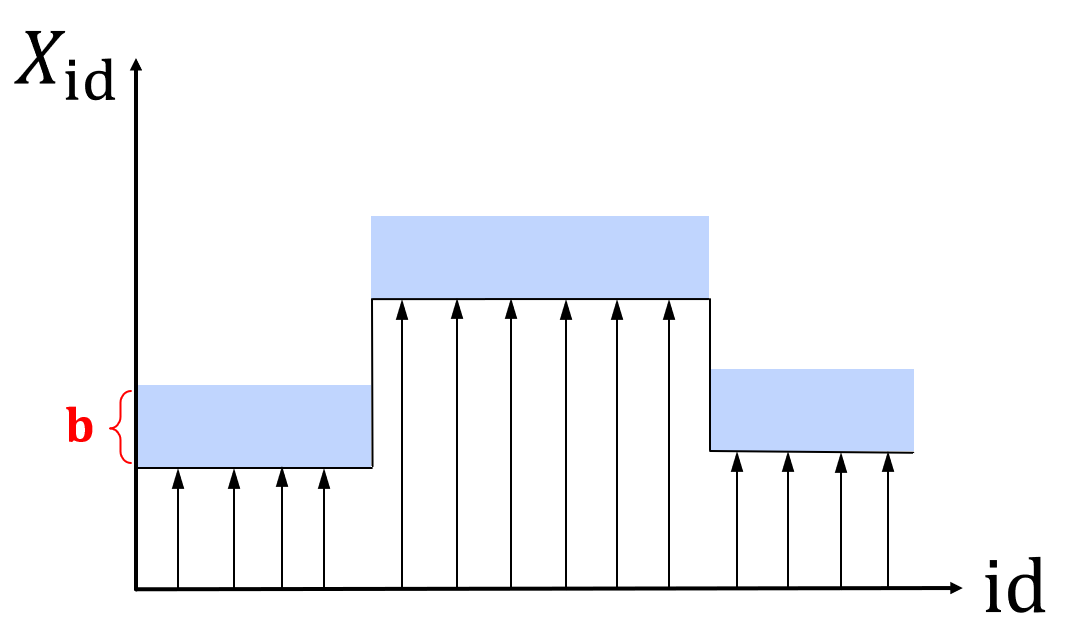}
      \caption{}
      \label{fig:data_raw}
    \end{subfigure}
    \hspace{-1em}
    \begin{subfigure}[b]{0.24\textwidth}
      \includegraphics[width=\textwidth]{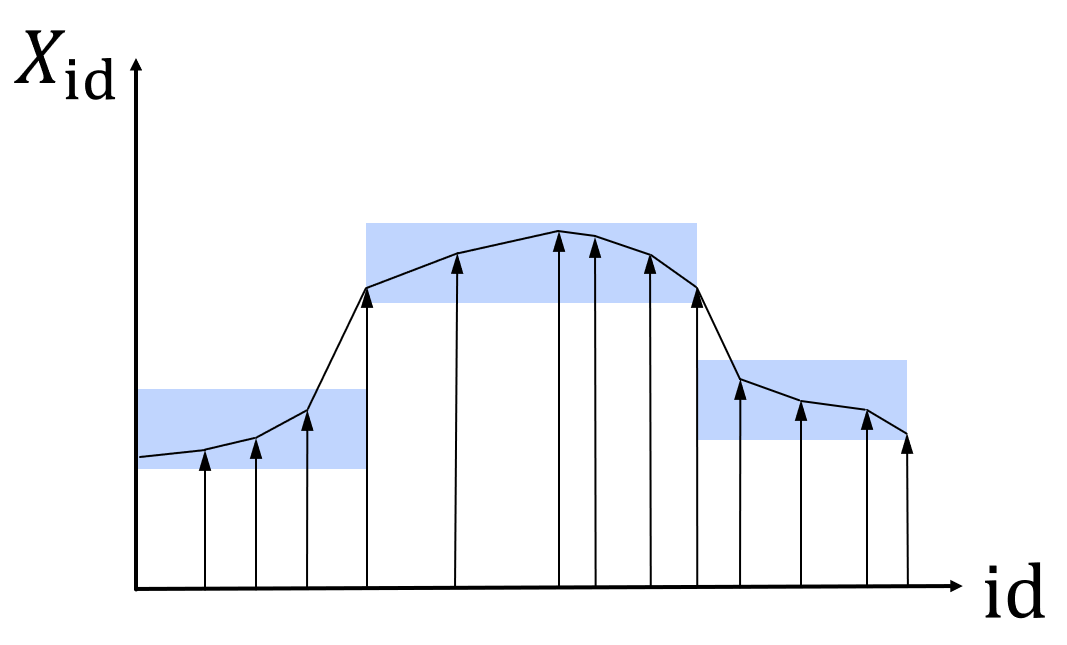}
      \caption{}
      \label{fig:data_dequantization}
    \end{subfigure}
    \caption{(a) Data before dequantization; (b) Data after dequantization.}
    \label{fig:dequantization}
  \end{figure}

\vspace{-4mm}
Consider a datapoint $\mathbf{x}$ and uniform noise $\mathbf{z} \in [0, \mathbf{b}]$, the dequantized data can be represented as $\mathbf x+\mathbf z$. For example, $Cate$ attribute has discrete values $\{1, 2, 3, 4, 5\}$ which can be more continuous like $\{1.347, 2.161, 3.481, 4.223, 5.439\}$ through dequantization, using a factor $b = 1$.

\noindent \textbf{Normalization} 
Since CUBE is trained on data with different dimensions, attributes with larger scales would have a greater effect to the others. We use z-score $normalization$ to scale the input features to a common scale, which helps to reduce the impact of outliers and improve the accuracy and generalization of the model. In the meantime, the gradient descent path would be smoother after normalization, leading to faster convergence. Let $\mathbf{x}$ be a datapoint from $T$, the normalized data with z-score can be represented as $(\mathbf x-\mathbf \mu) / \mathbf \delta$, where $\mathbf \mu$ is the mean value of the data and $\mathbf \delta$ is the standard deviation.

\subsection{Training Methods}

\subsubsection{\textbf{Unsupervised Learning using Table}} \mbox{} \\ 
After preprocessing table data with dequantization and normalization, we can train CUBE through unsupervised learning using. Our ultimate goal is to fit the joint distribution of the data as much as possible, i.e., to minimize the distance between the true data distribution $P(\mathbf x|\theta^{*})$ and the data distribution $P(\mathbf x|\theta)$ learned by the model. This distance can be measured using the $KL \ divergence$. According to the $Law \ of \ Large \ Numbers$, minimizing the $KL \ divergence$ is equivalent to maximizing the log-likelihood. Given a dataset $D = \{\mathbf x^{(i)}\}_{i = 1}^{N}$ consisting of $N$ datapoints on dimension $d$, we can use negative log-likelihood as loss function for unsupervised training to approximate the join data distribution.

\vspace{-4mm}
\begin{align}
\mathscr{L}_{data} 
&= -\frac{1}{N}\sum_{i=1}^N \log P(\mathbf x^{(i)}\|\theta) 
\end{align}

where $P(\mathbf x|\theta)$ can be obtained through our model\cite{gilboa2021mdma}:

\begin{equation}
\begin{aligned}
P(\mathbf x|\theta)  
&= \partial^d F(\mathbf x|\theta)/\partial \mathbf x \\ 
&= \sum\limits_{i_{1},\dots,i_{d}=1}^{m}A_{i_{1},\dots,i_{d}}\prod\limits_{j=1}^{d}\dot{\varphi_{i_{j},j}}(x_{j}|\theta).
\end{aligned}
\end{equation}

\subsubsection{\textbf{Supervised Learning using Workload}} \mbox{} \\ 
In data-driven cardinality estimation approaches, it is often assumed that the data distribution $f(\mathbf{X})$ is unknown. However, this assumption can be relaxed since our cardinality estimation model is intended to approximate the information of the tables in database.

Q-Error is widely used as a measure to evaluate cardinality estimators. Therefore, the model can directly minimize the average Q-Error of the training dataset as the optimization objective, where $Q_i$ denotes the $i$-th query, while $\widehat{s_i}$ and $s_i$ represent the estimation result and  true cardinality. The Q-Error is defined as below:

\begin{equation} \label{eq:qerror}
\text{Q-Error}(Q_i) = \frac{1}{|Q_i|} \sum_{i=1}^{|Q_i|} \max \left( \frac{s_i}{\widehat{s_i}}, \frac{\widehat{s_i}}{s_i} \right) 
\end{equation}

However, using Q-Error directly as a loss function can cause two problems, illustrated as below:
\begin{enumerate}
    \item The average Q-Error can be affected by a few large outliers (e.g., when the selectivity is 0, the Q-Error is unbounded). We address this issue by discarding excessively large Q-Errors.(e.g. 1e8). 
    \item The average Q-Error can fluctuate greatly during the training process, we use logarithmic function to map the Q-Error to a smaller and smoother range which helps to achieve stable training convergence. 
\end{enumerate}
    
Concurrently, we also notice that when Q-Error approaches 1, the corresponding value of the loss approaches 0, which does not help the model to converge well. Therefore, we add 1 to Q-Error before taking the logarithm to ensure the model can still converge well in the later stages of supervised training and learn the data distribution accurately. Thus, the query loss function is defined as below:
\begin{equation}
\mathscr{L}_{query} = \log(\text{Q-Error+1})
\end{equation}

\section{Experiments}
In this section, we first introduce the experimental setups, including the datasets, baselines, evaluation metrics, workload, hyper-parameter setting and environment. Then, we evaluate our model, CUBE, with extensive experiments to answer the following research questions:

\begin{itemize}[leftmargin=*]
    \item \textbf{RQ1:} How does CUBE compare to state-of-the-art cardinality estimators in accuracy and latency?
    \item \textbf{RQ2:} How significant is our inference acceleration, and how does the performance for CUBE  as dimension increases?
    \item \textbf{RQ3:} Does CUBE behave predicatable as theoretical analysis?
\end{itemize}

\subsection{Experimental Setup}\
\label{experimental_setup} 

\renewcommand{\descriptionlabel}[1]{\hspace{\labelsep}\normalfont #1}

\noindent \textbf{Datasets} We utilize real-world datasets, including BJAQ\cite{beijing_multi-site_air_quality_501}, Power\cite{individual_household_electric_power_consumption_235}, and YearPredictMSD\cite{year_prediction_msd_203} as single tables from the UCI Machine Learning repository. Additionally, we use the multi-table Employees dataset\cite{employees} from commercial MySQL databases. These datasets feature complex correlations between columns, enabling a more comprehensive comparison of model performance.

\noindent \textbf{Baselines} 
We include these cardinality estimators for experimental comparison. 

\begin{description}
  \item[1) PG] Postgres's internal method which using histograms with the independence assumption.
  \item[2) Sample\cite{leis2017cardinality,zhao2018random}] a method sampled a certain piece of records to perform cardinality estimation. 
  \item[3) lw-nn\cite{dutt2019selectivity}] a query-driven method that trained a neural network for cardinality estimation.
  \item[4) MSCN\cite{kipf2018learned}] a query-driven method that used multi-set convolutional network.
  \item[5) PRICE\cite{zeng2024price}] a query-driven method that pretrained on 25 diverse datasets without Employee.
  \item[6) DeepDB\cite{hilprecht13deepdb}] a data-driven method used sum-product network.
  \item[7) Naru\cite{yang13deep}/NeuroCard\cite{yang2020neurocard}] a data-driven method used the autoregressive model.
  \item[8) FACE\cite{wang2024face}] a data-driven method used normalizing flow. 
  \item[9) CUBE-d/CUBE-q] Our CardEst method based on CDF, the model trained using table is denoted as CUBE-d, while model trained using the workload is denoted as CUBE-q.
  
\end{description} 
\noindent \textbf{Evaluation Metrics} We comprehensively evaluate CardEst models from the following perspectives: (1) Performance, including accuracy using Q-error, inference latency and End-to-End(E2E) latency (2) Scalability, which reflects the model’s ability to maintain strong performance as data dimensionality grows (3) Predictability, including fundamental logical rules (monotonicity, validity, consistency and stability) introduced in Section ~\ref{subsec:predictable_analysis}.

\noindent \textbf{Workload} 
For each dataset, we generated 2,000 test queries included both range and equality predicates with their true cardinalities, which were generated by: (a) randomly selecting the number of predicates $k$ within a reasonable range based on the dataset's number of columns; (b) randomly selecting $k$ distinct columns to place the predicates. For numerical columns, randomly choosing the predicate type from $\{ =, \leq, \geq \}$ ; for categorical columns, only generating equality predicates as range predicates were treated impractical; (c) randomly selecting a tuple from the table and using its attribute values as the predicate literals.

\noindent \textbf{Hyper-parameter Setting} 
For our model, $m$ is the width of CUBE, $l$ and $r$ are respectively the depth and width of the univariate CDF model. We trained with $m$ = 1000, $l$ = 2, $r$ = 3 for 1000 epochs and learning rate 0.01 on all datasets. Then, we configured other baseline models with their default hyper-parameter settings.

\noindent \textbf{Environment}
All experiments were in Python, performed on a server with Intel(R) Xeon(R) Silver 4310 CPU \@ 2.10GHz, Nvidia A5000 GPU, and 251GB RAM.
\begin{table}[h!]
\centering
\caption{BJAQ dataset evaluation results}
\begin{tabular}{l|cccc|c}
    \toprule
    \textbf{Estimator} & 50th & 95th & 99th & Max & InferLat(ms)\\
    \midrule
    PG          & 1.7  & 10.78 & 40.4  & 1390 & \textbf{0.31} \\
    Sample      & 1.10 & 1.41  & 2.23  & 315  & 1.12  \\
    lw-nn       & 1.13 & 6.06  & 15.7  & 80.9 & 0.91  \\
    MSCN        & 1.18 & 2.45  & 11.0  & 178  & 1.05  \\
    DeepDB      & 1.09 & 1.93  & 5.57  & 492  & 3.91  \\
    Naru        & 1.03 & 1.27  & 1.71  & 9.31 & 10.4  \\
    FACE        & 1.03 & 1.15  & 1.19  & 1.86 & 9.7   \\
    *CUBE-d     & \textbf{1.01} & \textbf{1.07} & \textbf{1.17} & \textbf{1.60} & 0.60  \\
    *CUBE-q     & 1.01 & 1.08  & 1.24  & 10.00  & 0.66  \\
    \bottomrule
\end{tabular}
\label{tab:bjaq}
\end{table}
    
\begin{table}[h!]
\centering
\caption{Power dataset evaluation results}
\begin{tabular}{l |cccc |c}
    \toprule
    \textbf{Estimator} & 50th & 95th & 99th & Max & InferLat(ms)\\
    \midrule
    PG          & 1.32 & 17.1  & 121   & $3\text{e}^5$ & 1.13  \\
    Sample      & 1.09 & 2.01  & 167   & 735   & 2.13  \\
    lw-nn       & 1.06 & 4.83  & 27.0  & 460   & \textbf{0.51}  \\
    MSCN        & 1.15 & 17.7  & 181   & 510   & 0.69  \\
    DeepDB      & 1.07 & 1.89  & 5.45  & 551   & 15.40 \\
    NeuroCard   & 1.02 & 1.53  & 5.11  & 123   & 70.9    \\
    FACE        & 1.03 & 1.20  & 1.75  & 5.00  & 8.78  \\
    *CUBE-d     & 1.02 & \textbf{1.10} & \textbf{1.34} & \textbf{2.80} & 0.60 \\
    *CUBE-q     & \textbf{1.01} & 1.19  & 1.91  & 69.00   & 0.62  \\
    \bottomrule
\end{tabular}
\label{tab:power}
\end{table}
    
\begin{table}[h!]
\centering
\caption{Employees dataset evaluation results}
\setlength{\tabcolsep}{2.5pt} 
\begin{tabular}{l |cccc |c}
    \toprule
    \textbf{Estimator} & 50th & 95th & 99th & Max & InferLat(ms)\\
    \midrule
    PG          & 38.03  & 1963.12 & 8378.8  & $2\text{e}^5$ & \textbf{0.55} \\
    MSCN        & 1.08 & \textbf{1.50}  & 2.71  & 8.6  & 0.67  \\
    PRICE       & 15.90 & 123.96 & 211.17 & 318.50 & 0.75 \\
    PRICE(finetune) & 1.73 & 6.19 & 15.47 & 86.13 & 0.698 \\
    DeepDB      & \textbf{1.075} & 2.38  & 4.87  & 13.64  & 13.01  \\
    NeuroCard   & 1.11 & 2.05 & 3.37 & 14.47 & 69.0  \\
    *CUBE-d     & 1.099 & 1.61 & \textbf{2.51} & \textbf{4.31} & 2.4  \\
    \bottomrule
\end{tabular}
\label{tab:employees}
\end{table}

\begin{figure}[h]
  \centering
  \begin{subfigure}[b]{0.21\textwidth}
    \includegraphics[width=\textwidth]{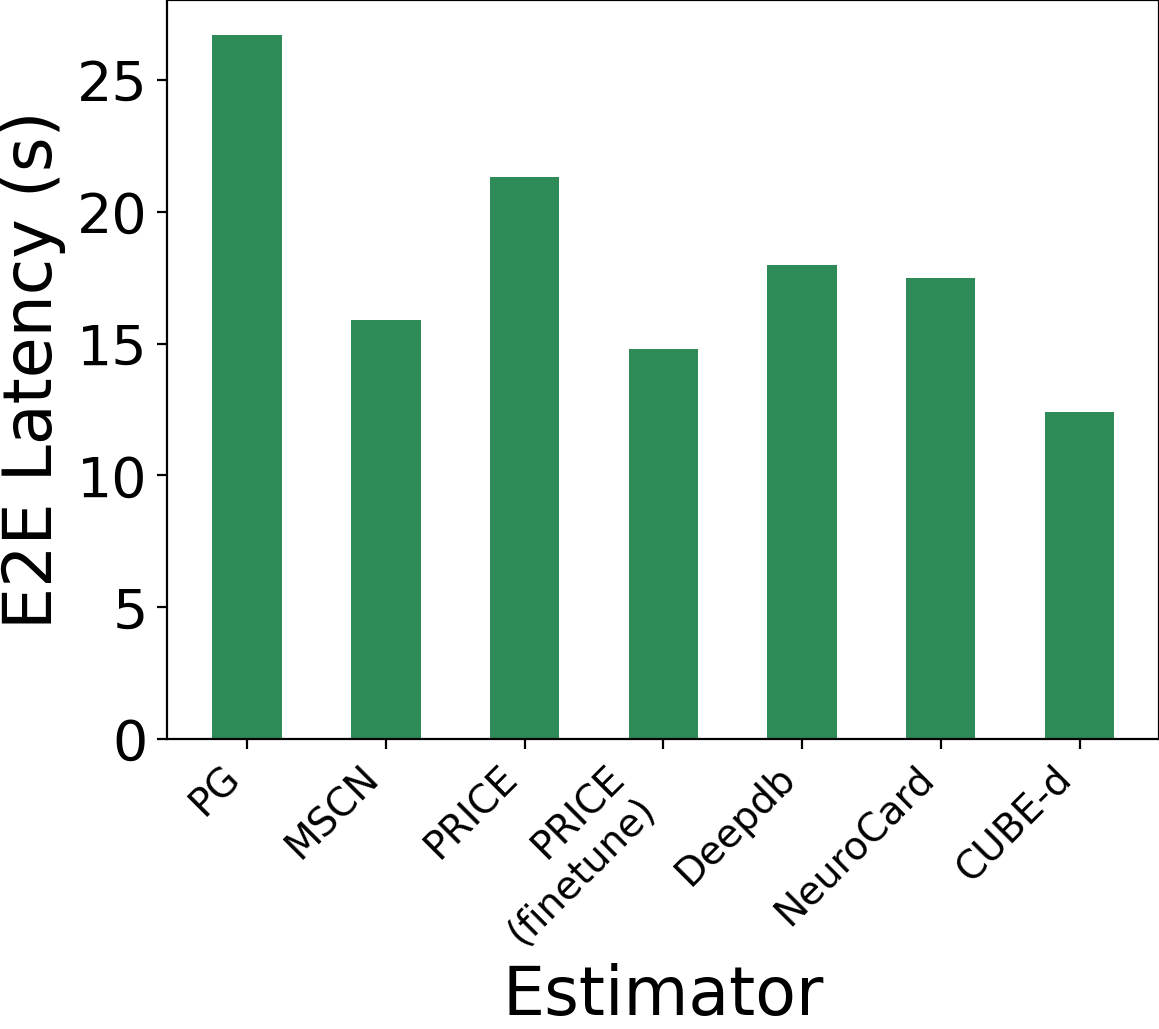}
    \caption{}
    \label{fig:e2e_time}
  \end{subfigure}%
  \hspace{3mm}
  \begin{subfigure}[b]{0.24\textwidth}
    \includegraphics[width=\textwidth]{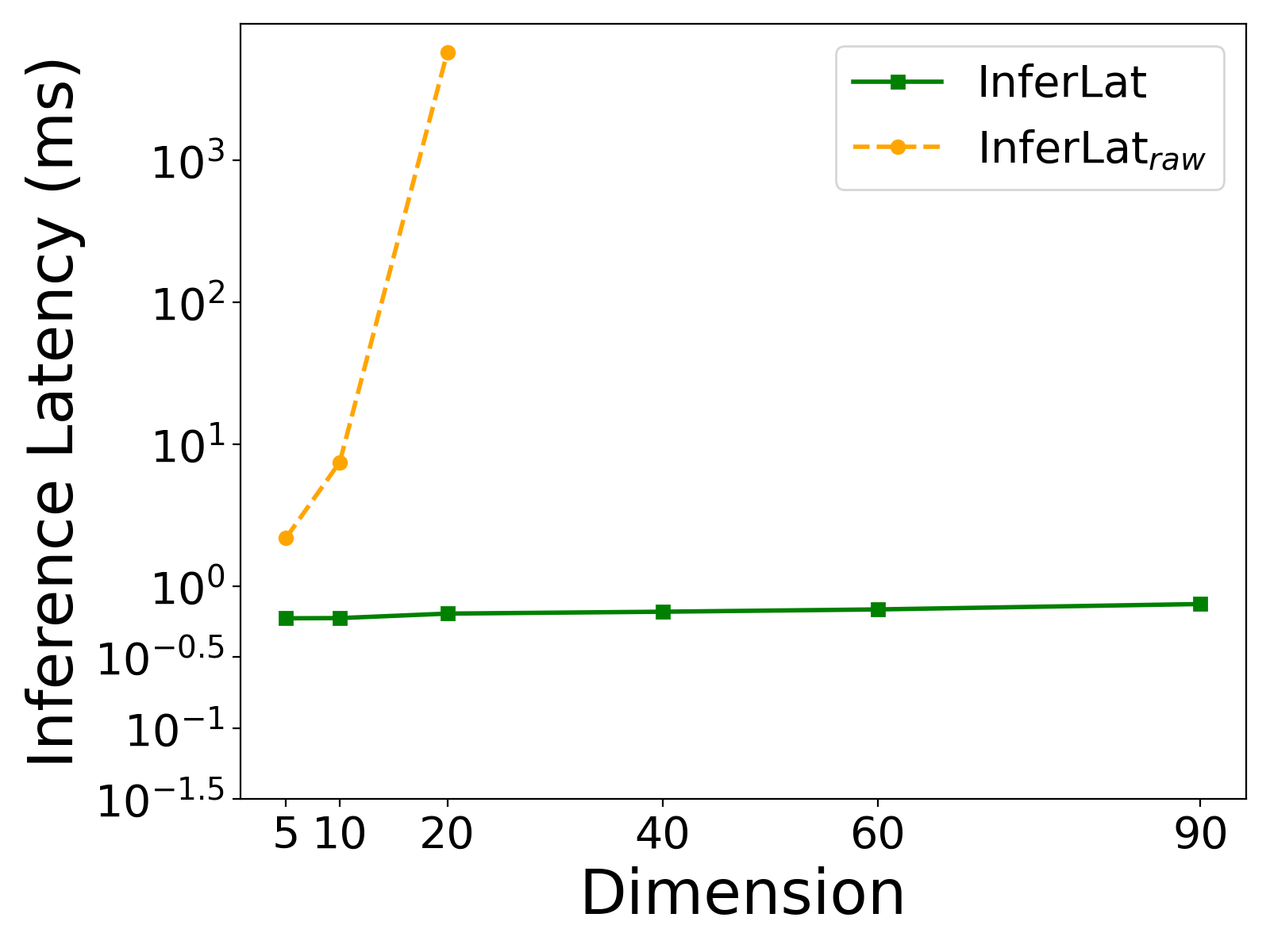}
    \caption{}
    \label{fig:infer_time}
  \end{subfigure}
  \caption{(a) E2E latency(s) on Employee dataset; (b) CUBE's log-scale latency (ms) on YearPredictMSD dataset.}
  \label{fig:latency}
\end{figure}

\subsection{Performance Evaluation} 
In this section, we compare the performance of CUBE based on cumulative distribution with other methods. In single-table CardEst scenario, we report the evaluation reslut on BJAQ dataset in Table ~\ref{tab:bjaq} and Power dataset in Table ~\ref{tab:power}. Due to the large domain size of the Power dataset, Naru could not be effectively trained and we only report the results for NeuroCard; While BJAQ dataset has a medium domain size, we report Naru's results because Naru and NeuroCard employ the same underlying methodology. 

For the multi-table CardEst scenario, we selected 4 tables including \textit{salaries}, \textit{dept\_emp}, \textit{employee}, and \textit{titles} from Employees dataset for evaluation shown in Table ~\ref{tab:employees}. lw-nn does not support join predicate, and FACE's code repository does not provide code for multi-table CardEst. CUBE-d share the same architecture as CUBE-q and has better performance, thus we conducted comparative experiments for multi-table joins with CUBE-d, PG, MSCN, PRICE, DeepDB and NeuroCard.

\noindent \textbf{Accuracy} CUBE-d achieve the highest accuracy among these methods and outperforms on BJAQ and Pozwer datasets, even beaten Naru and FACE. CUBE-d performs particularly well on tail Q-error on all datasets, the 95th q-error (1.07, 1.10 and 1.61) is very close to 1 and max q-error in all cases is smaller than 5. Under sinlgle-table CardEst scenario, their accuracy can be ranked as CUBE-d $\approx$ FACE > NARU/NeuroCard > CUBE-q > DeepDB $\gg$ lw-nn/MSCN/Sample > PG; under muti-table scenario, the ranking is CUBE-d  > MSCN >  DeepDB/NeuroCard > PRICE $\gg$ PG. PG has large errors due to its assumption of attribute independence, which leads to lower accuracy when attributes are strongly correlated. Sampling method performs well on median metrics across datasets but shows significantly lower accuracy on tail metrics for handling low-selectivity queries poorly.

The accuracy of FACE and Naru was already very high, but they incur errors not only from the model but also from Monte Carlo integration and progressive sampling, which further amplifies the estimation errors. Moreover, NeuroCard uses column factorization to handle large domains, but this approach comes at the cost of reduced accuracy. DeepDB performs well on median metrics but shows significantly tail errors when attributes are highly correlated. This is primarily due to DeepDB making independence assumptions in some SPN nodes.

CUBE-d also achieves higher accuracy than other query-driven methods. It shares the same architecture as CUBE-q and leverages workload information to better fit the data distribution. Thus, it can provide more accurate predictions compared to unconstrained, regression-based query-driven methods. MSCN can handle attributes with large domains and strong correlations relatively well which help MSCN perform better than Deepdb/NeuroCard on Employees dataset. But MSCN still has errors reaching 500 in tail scenario on Power dataset, which is possibly because MSCN heavily relies on the distribution of training queries, making low-selectivity data queries rare. PRICE achieves practical accuracy without fine-tuning, which highlighting its strong generalization capabilities. The accuracy of PRICE can be further improved after fine-tuning, but still lower than other learning-based methods.

\noindent \textbf{Inference Latency} Traditional methods (e.g. PG based on one-dimensional histograms) were very fast because they were very simple, but suffered from low accuracy. After inference acceleration, CUBE-d achieve low latency within 1ms under single-table scenario while maintaining high accuracy, whereas FACE and Naru were 10x slower than CUBE. FACE involves high-dimensional integration over probability densities to get result, which is significantly more time-consuming than the inference process of the NF model; Naru/NeuroCard relies on sampling for range query results, introducing additional overhead. Particularly for datasets with large domains (e.g. Power), NeuroCard’s column factorization increases the number of columns requiring sampling, further amplifying computational overhead and thus leading to higher inference latency. 

We can clearly observe that in multi-table CardEst scenario, the inference latency of CUBE is 4x slower than that of single-table CardEst. This is because CUBE needs to calculate the conditional probability for each equivalent key variable. However, it is still faster than other data-driven methods.

In contrast, regression models do not require sampling of range predicates and perform inference using only encoded queries, thus having higher efficiency and lower latency. Additionally, CUBE-d's inference latency is comparable to or even lower than other regression-based cardinality estimation methods (e.g. MSCN). Under the same parameters, CUBE-d and CUBE-q have similar latencies due to their shared model architecture.

\noindent \textbf{E2E Latency}
We evaluate E2E latency on Employees dataset with different estimators by inject estimation results into modified PG, which is the ultimate goal of optimizing CardEst within a query optimizer. Typically, query optimizer could generate better query plans with more accurate estimation results. As shown in Figure \ref{fig:e2e_time}, CUBE achieves the shortest execution time, since our model provides the most accurate estimation. Compared to the naive CardEst method used in PG, CUBE reduces the execution time by 50\%. However, we observed that the above situation is not always satisfied. Different estimation results might generate similar query plans. Although the CardEst results of PRICE after fine-tuning are not as accurate as DeepDB/NeuroCard, it can still generate better execution plans.

\subsection{Scalability Evaluation}
For evaluating the scalability of CUBE as dimension increases, we performed experiments using subsets of the YearPredictMSD data with 5, 10, 20, 40, 60, and 90 columns as shown in Table \ref{tab:yearpredict}. Since CUBE-d and CUBE-q share the same architecture and CUBE-d exhibits  higher accuracy, we focus on presenting the results on CUBE-d. The log-scale latency (comparing cases with and without inference acceleration) as dimensionality increases is visualized in Figure.~\ref{fig:infer_time}.

The inference latency with optimization (InferLat) remains almost unchanged as dimensionality increases. In contrast, the latency without inference acceleration ($\text{InferLat}_{raw}$) is 10x higher than the optimized version at a dimensionality of 10 and exhibits exponential growth as the dimensionality increases. We observe that the tail estimation increases very slowly with dimensionality, remaining within 100 even at 90 dimensions. Meanwhile, the training time (TT) grows linearly, demonstrating the model's excellent scalability with respect to dimensionality.

\begin{table}[h!]
    \centering
    \setlength{\tabcolsep}{1.5pt} 
    \caption{Evaluation result of CUBE on YearPredictMSD}
    \begin{tabular}{l|cccc|cc|c}
        \toprule
        \textbf{Dim} & 
        50th & 95th & 99th &Max & 
        $\text{InferLat}_{raw}$(ms)&  InferLat(ms) &
        TT(h)\\
        \midrule
        5   & 1.01 & 1.05  & 1.17  & 1.70   & 2.19 & 0.595  & 0.25  \\
        10  & 1.03 & 1.28  & 1.62  & 6.00   & 7.49 & 0.597 & 0.35 \\
        20  & 1.14 & 3.00  & 4.67  & 9.00   & 5786.88 & 0.642 & 0.80 \\
        40  & 1.17 & 3.00  & 5.00  & 14.50  & - & 0.662 & 1.72 \\
        60  & 1.29 & 5.00  & 12.00 & 73.00  & - & 0.687 & 2.80 \\
        90  & 2.00 & 27.00 & 87.07 & 277.25 & - & 0.750 & 4.99 \\
        \bottomrule
    \end{tabular}
    \label{tab:yearpredict} 
\end{table}

\subsection{Predictability Evaluation}
According to the rules proposed in Section~\ref{subsec:predictable_analysis}, we evaluates learned cardinality estimation models and summarizes whether they satisfy or violate each rule in Table~\ref{tab:predictivity}. 

\begin{figure}[htp]
  \includegraphics[width=0.45\textwidth]{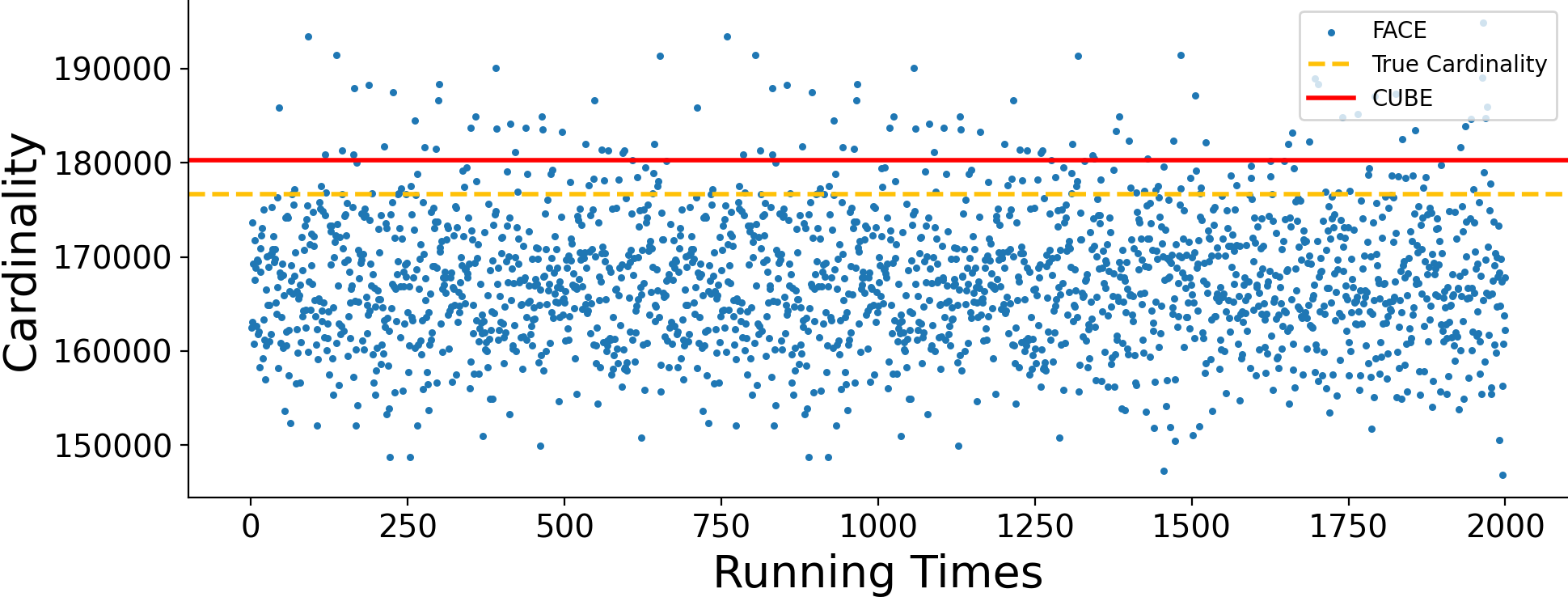}
  \caption{Case study of stability comparing FACE with CUBE}
  \label{fig:stability}
\end{figure}

Although Naru/NeuroCard and FACE maintain high accuracy, they lose stability. The progressive sampling technique of Naru/NeuroCard and the Monte Carlo integration process of FACE introduce uncertainty during inference, violating the stability rule. Currently, no method achieves both high accuracy and high stability simultaneously. However, CUBE based on CDF ensures the highest accuracy while maintaining stability. Figure~\ref{fig:stability} illustrates an example of estimation results from running the same query 2000 times using the FACE and CUBE models.

This instability also causes Naru/NeuroCard and FACE to violate monotonicity and consistency rules. Regression-based methods (MSCN, lw-nn) violate all rules except stability, as no constraints are applied to the models during training or inference.

In contrast, CUBE based on CDF construction and DeepDB based on SPN construction do not violate any rules. The former satisfies the probability characteristics of CDF by applying constraints to neural networks, and Section~\ref{subsec:predictable_analysis} demonstrates that CUBE meets the four fundamental probability rules. The latter is constructed based on basic histograms, and calculations between nodes are limited to addition and multiplication. Because some learned cardinality estimation methods fail to adhere to fundamental logical rules, database systems may exhibit illogical behaviors, confusing users. For example, users typically expect queries to run faster with additional filtering conditions, but when using models like FACE, Naru/NeuroCard, MSCN, or lw-nn, the opposite situation may occur due to violations of the monotonicity rule.

\newcolumntype{g}{>{\columncolor{green!15}}c}
\begin{table}[h]
\centering
\caption{Basic logical rules satisfied situation}
\setlength{\tabcolsep}{1.5pt} 
\begin{tabular}{l |c c c g c c g} 
\hline
\textbf{Rules} & lw-nn & MSCN & PRICE & \multicolumn{1}{c}{\cellcolor{white}DeepDB} & Naru & FACE & \multicolumn{1}{c}{\cellcolor{white}CUBE} \\
\hline
Monotonicity & \(\times\) & \(\times\) & \(\times\) & \(\checkmark\) & \(\times\) & \(\times\) & \(\checkmark\) \\
Validity   & \(\checkmark\) & \(\checkmark\) & \(\checkmark\) & \(\checkmark\) & \(\checkmark\) & \(\checkmark\) & \(\checkmark\) \\
Consistency  & \(\times\) & \(\times\) & \(\times\) & \(\checkmark\) & \(\times\) & \(\times\) & \(\checkmark\) \\
Stability    & \(\checkmark\) & \(\checkmark\) & \(\checkmark\) & \(\checkmark\) & \(\times\) & \(\times\) & \(\checkmark\) \\
\hline
\end{tabular}
\vspace{3mm}
\label{tab:predictivity}
\end{table}

\vspace{-4mm}

\vspace{-4mm}
\section{Related Work}

\noindent \textbf{Density Estimation}
Estimating the joint probability density of random variables is a fundamental task in statistics, which can capture the distribution of given data. Traditional density estimators such as histograms and kernel density estimators(KDEs) \cite{heimel2015self,park2020quicksel} typically perform well only in low dimension. Neural network-based approaches were proposed for density estimation, and yielded promising results for high dimensional problems. There are mainly two families of such neural density estimators: autoregressive models\cite{germain2015made,vaswani2017attention,uria2013rnade,uria2013rnade,gregor2014deep} and normalizing flows\cite{dinh2014nice,rezende2015variational}. Deep generative models such as Roundtrip\cite{liu2020roundtrip} are also adopted to density estimation task later.

\noindent \textbf{Learned Cardinality Estimation} Machine learning based techniques have been proposed to solve CardEst problem, which can achieve accurate estimation results. Specifically, the query-driven methods learn a regression model from a query to its cardinality, such as MSCN\cite{kipf2018learned}, ALECE\cite{li2023alece} and LPCE\cite{wang2023speeding}. While data-driven methods learn a distribution of tuples, which can be used to estimate query's cardinality. Naru\cite{yang13deep}/NeuroCard\cite{yang2020neurocard} used deep autoregressive models, FACE utilizes normalizing flow models, and DeepDB\cite{hilprecht13deepdb}/FLAT\cite{zhu2020flat} adopts sum-product networks\cite{desana2020sum,poon2011sum,molina2018mixed,gens2013learning} to perform inference over joint distribution to produce estimation results. To support multi-table CardEst with join predicate, one can either build a single global model, like NeuroCard, or construct multiple local models by partitioning table schema, such as DeepDB and Scardina\cite{ito2023scardina}. Factorjoin\cite{wu2023factorjoin} propose a framework by performing inference over the factor graph involving multiple individual models built for single-table distributions.

\section{Conclusion and Future Work}
We have shown that CardEst using CDF can achieve accurate and predictable results, which guarantees the consistency of generated execution plans, contributing to the consistent performance of commercial databases. Our experiments demonstrate that the inference acceleration strategy for high-dimensional data significantly reduces latency, resulting in a substantial performance improvement. CUBE has tail Q-error close to 1 with 10x faster inference speed than current SOTA data-diven methods. With low inference time, our model exhibits excellent performance on dimensional scalability, proving particularly valuable for large-scale data processing.

For future work, hybrid training using both table and workload is an interesting direction, which may achieve higher accuracy and shorter training time. Furthermore, by integrating our model as a component into GPU databases, we may unlock even greater performance improvements, enabling faster query processing.




\clearpage
\appendix
\section{Proof of Corollary 1} \label{subsec:proof_corollary_1}

\begin{corollarynum}{\ref{corollary:inference_optimized}}

Given random variable $\mathbf{X} = (X_1, X_2, \dots, X_d)$ with CDF approximation as $\sum\limits_{i_{1},\dots,i_{d}=1}^{m}A_{i_{1},\dots,i_{d}}\prod\limits_{j=1}^{d}\varphi_{i_{j},j}(x_{j})$ and a query region $\Omega = [l_1, u_1] \times \cdots \times [l_d, u_d]$. The probability that $\mathbf{x}$ falls in $\Omega$ is 

\begin{equation*}
\begin{split}
\Pr(\mathbf x \in \Omega)
&= \sum_{\mathbf{s}\in\{0,1\}^n} (-1)^{n-|\mathbf{s}|}\, F(\mathbf{v})\\
&\approx
    \sum_{i_{1},\dots,i_{d}=1}^{m}
      A_{i_{1},\dots,i_{d}}
      \prod_{j=1}^{d}
        \bigl[\varphi_{i_{j},j}(u_{j}) - \varphi_{i_{j},j}(l_{j})\bigr].
\end{split}
\end{equation*}

\end{corollarynum}

\begin{proofnum} 
When d = 2, random variable $\mathbf{X} = (X_1, X_2)$, the probability that $\mathbf{x}$ falls into the region $\Omega=[l_1,u_1]\times [l_2, u_2]$ is:
\begin{align*}
&\Pr(l_1\leq x_1 < u_1,\, l_2\leq x_2 < u_2) \\
&= F(u_1, u_2) - F(l_1, u_2) - F(u_1, l_2) + F(l_1, l_2) \\  
&\approx \Biggl[\sum_{i,j=1}^{m} A_{i,j}\varphi_{i,1}(u_{1})\varphi_{j,2}(u_{2})
- \sum_{i,j=1}^{m} A_{i,j}\varphi_{i,1}(l_{1})\varphi_{j,2}(u_{2})\Biggr] \\  
&\quad - \Biggl[\sum_{i,j=1}^{m} A_{i,j}\varphi_{i,1}(u_{1})\varphi_{j,2}(l_{2})
- \sum_{i,j=1}^{m} A_{i,j}\varphi_{i,1}(l_{1})\varphi_{j,2}(l_{2})\Biggr] \\  
&= \sum_{i,j=1}^{m} A_{i,j}\Bigl[\varphi_{i,1}(u_{1}) - \varphi_{i,1}(l_{1})\Bigr]\varphi_{j,2}(u_{2}) \\  
&\quad - \sum_{i,j=1}^{m} A_{i,j}\Bigl[\varphi_{i,1}(u_{1}) - \varphi_{i,1}(l_{1})\Bigr]\varphi_{j,2}(l_{2}) \\  
&= \sum_{i,j=1}^{m} A_{i,j}\Bigl[\varphi_{i,1}(u_{1}) - \varphi_{i,1}(l_{1})\Bigr]
\Bigl[\varphi_{j,2}(u_{2}) - \varphi_{j,2}(l_{2})\Bigr]
\end{align*}

The general case can be proved by mathematical induction. 
\end{proofnum}
        
\section{Proof of Corollary 2} \label{subsec:proof_corollary_2}

\begin{corollarynum} {\ref{corollary:inference_expectation}}

Given random variable $X = (X^{(1)}, X^{(2)})$ with CDF approximation as
$\sum\limits_{i_{1},\dots,i_{d}=1}^{m}A_{i_{1},\dots,i_{d}}\prod\limits_{j=1}^{d}\varphi_{i_{j},j}(x_{j})$, where random vectors $X^{(1)} = (X_1, \ldots, X_k)$ and $X^{(2)} = (X_{k+1}, \ldots, X_d)$ represent two sub-vectors of $X$. Let $W(X^{(1)})$ be a function depending only on sub-vector $X^{(1)}$, with domain $R_{X}^{(1)}$ for $X^{(1)}$ and domain $\Omega = [l_{k+1}, u_{k+1}] \times \dots \times [l_{d}, u_{d}]$ for $X^{(2)}$. Then the expectation of $W(X^{(1)})$ within region $\Omega$ is present as:

\begin{equation*}
\begin{aligned} 
&\mathbb{E}\left[\mathds{1}_{\{x^{(2)} \in \Omega\}}\cdot W(x^{(1)})\right] \\
&= \sum_{i_{1},\dots,i_{d}=1}^{m}A_{i_{1},\dots,i_{d}}\left\{\prod\limits_{j=k+1}^{d}[\varphi_{i_{j},j}(u_{j})-\varphi_{i_{j},j}(l_{j})]\right\}\cdot G_{i_1,\dots,i_k}(x^{(1)}).
\end{aligned}
\end{equation*}

where intermediate function $G_{i_1,\dots,i_k}(x^{(1)})$ is defined as:

\begin{equation*}
\begin{aligned} 
&G_{i_1,\dots,i_k}(x^{(1)}) \\
&= \sum_{x^{(1)} \in R_{X}^{(1)}}\frac{\prod\limits_{j=1}^{k}\dot{\varphi}_{i_{j},j}(x_{j})}{\sum\limits_{i_{1},\dots,i_{k}=1}^{m}A_{i_{1},\dots,i_{k}}\prod\limits_{j=1}^{k}\dot{\varphi}_{i_{j},j}(x_{j})}\cdot\Pr\left(X^{(1)}=x^{(1)}\right)\cdot W(x^{(1)}).
\end{aligned}
\end{equation*}

\end{corollarynum}

\begin{proofnum}
Consider the random variable $X = (X^{(1)}, X^{(2)})$. The expectation of $W(X^{(1)})$ within region $\Omega$ can be expressed as:
\begin{align*} 
&\mathbb{E}\left[\mathds{1}_{\{x^{(2)} \in \Omega\}}\cdot W(x^{(1)})\right] \\
&= \sum_{x^{(1)} \in R_{X}^{(1)}}\Pr\left(x^{(2)}\in\Omega, X^{(1)}=x^{(1)}\right)\cdot W(x^{(1)}) \\
&= \sum_{x^{(1)} \in R_{X}^{(1)}}\Pr\left(x^{(2)}\in\Omega \mid x^{(1)}\right)\cdot\Pr\left(X^{(1)}=x^{(1)}\right)\cdot W(x^{(1)}).
\end{align*}

Similar to Theorem~\ref{thm:inference_raw}, the conditional probability of $X^{(2)}$ within region $\Omega$ can be expressed by conditional CDF:
\begin{align*} 
\Pr\left(x^{(2)}\in\Omega \mid x^{(1)}\right) = \sum_{\mathbf{s}\in\{0,1\}^{d-k}}(-1)^{\,d-k-|\mathbf{s}|}\,F_{x^{(2)}|x^{(1)}}(\mathbf{v}_{\mathbf{s}}).
\end{align*}

The conditional CDF approximation using MDMA derivatives is:
\begin{align*}
\widehat{F}(x_{k+1},\dots,x_{d}\mid x_{1},\dots,x_{k}) 
= \frac{\sum\limits_{i_{1},\dots,i_{d}=1}^{m}
A_{i_{1},\dots,i_{d}}
\prod\limits_{j=k+1}^{d}\varphi_{i_{j},j}(x_{j})
\prod\limits_{j=1}^{k}\dot{\varphi}_{i_{j},j}(x_{j})
}{
\sum\limits_{i_{1},\dots,i_{d}=1}^{m}
A_{i_{1},\dots,i_{d}}
\prod\limits_{j=1}^{k}\dot{\varphi}_{i_{j},j}(x_{j})
}.
\end{align*}

Similar to Corollary~\ref{corollary:inference_optimized}, the conditional probability of $X^{(2)}$ within region $\Omega$ can be simplified by combining calculations:
\begin{align*} 
&\Pr\left(x^{(2)}\in\Omega \mid x^{(1)}\right) \\ 
&= \frac{\sum\limits_{i_{1},\dots,i_{d}=1}^{m}A_{i_{1},\dots,i_{d}}\left\{\prod\limits_{j=k+1}^{d}[\varphi_{i_{j},j}(u_{j})-\varphi_{i_{j},j}(l_{j})]\right\}\prod\limits_{j=1}^{k}\dot{\varphi}_{i_{j},j}(x_{j})}{\sum\limits_{i_{1},\dots,i_{k}=1}^{m}A_{i_{1},\dots,i_{k}}\prod\limits_{j=1}^{k}\dot{\varphi}_{i_{j},j}(x_{j})}.
\end{align*}

Substituting back into the expectation expression yields the final result:
\begin{align*}
&\mathbb{E}[\mathds{1}_{\{x^{(2)} \in \Omega\}}W(x^{(1)})] \\
&= \sum_{i_1,\dots,i_d=1}^{m} A_{i_1,\dots,i_d}
\Big\{\prod_{j=k+1}^{d}[\varphi_{i_j,j}(u_j)-\varphi_{i_j,j}(l_j)]\Big\}
\cdot G_{i_1,\dots,i_k}(x^{(1)}).
\end{align*}
\end{proofnum}

\clearpage
\bibliographystyle{ACM-Reference-Format}
\bibliography{reference}

\appendix
\end{sloppypar}
\end{document}